\documentclass[sigconf]{acmart}

\usepackage{sigirstyle}
\usepackage{appendix}
\usepackage{balance}

\usepackage{amsmath}
\usepackage{booktabs}
\usepackage{tabularx}
\usepackage{caption}
\usepackage{subcaption}
\usepackage{adjustbox}
\usepackage{lineno}
\usepackage{makecell}

\usepackage[utf8]{inputenc}
\usepackage{url}

\usepackage{pgfplots}
\pgfplotsset{compat=1.18}
\usepackage{pgfplotstable}
\definecolor{methodA}{HTML}{6BAED6}  
\definecolor{methodB}{HTML}{2CA25F}  
\definecolor{methodC}{HTML}{CC6633}  
\definecolor{methodD}{HTML}{9E6B8A}  

\usetikzlibrary{calc,shapes.geometric}
\usepackage{xcolor}
\definecolor{cLow}{HTML}{4C9BE8}
\definecolor{cMed}{HTML}{F0A500}
\definecolor{cHigh}{HTML}{2EAA6E}
\definecolor{cXH}{HTML}{E05A5A}
\definecolor{cQPP}{HTML}{8B2FC9}
\definecolor{cSFT}{HTML}{E8500A}
\definecolor{axisbg}{HTML}{F8F9FA}

\usepackage{pgfplots}
\pgfplotsset{compat=1.18}
\usepackage{pgfplotstable}
\usetikzlibrary{pgfplots.groupplots}

\usepackage{calrsfs}

\DeclareMathAlphabet{\pazocal}{OMS}{zplm}{m}{n}
\DeclareMathAlphabet{\pazobfcal}{OMS}{cmsy}{b}{n}

\definecolor{lightblue}{RGB}{173, 216, 230}

\definecolor{mygray}{gray}{0.9} 

\DeclareMathOperator*{\argmin}{arg\,min}
\newcommand\mybox[2][]{\tikz[overlay]\node[fill=blue!20,inner sep=2pt, anchor=text, rectangle, rounded corners=1mm,#1] {#2};\phantom{#2}}

\newcommand{\gbox}[1]{\mybox[fill=green!10]{#1}}

\newcommand{\rbox}[1]{\mybox[fill=red!10]{#1}}

\newcommand{\qpprag}{DRAG$_\text{QPP}$}
\newcommand{\adarag}{DRAG}
\newcommand{\sOracle}{RAG$_\text{Orac}$}
\newcommand{\wOracle}{DRAG$_\text{AP}$}
\newcommand{\iclnz}{0-shot}
\newcommand{\static}{RAG$_\text{max}$}
\newcommand{\sft}{DRAG$_\text{SFT}$}
\newcommand{\configZero}{DRAG$_\text{PPT}$}

\newcommand{\DG}[1]{}
\newcommand{\PS}[1]{}
\newcommand{\NA}[1]{}
\newcommand{\SB}[1]{}
\newcommand{\PBC}[1]{}

\usepackage{tikz}
\usepackage{booktabs}      
\usepackage{multirow}      
\usepackage{colortbl}      
\usepackage{xcolor}        
\usepackage{adjustbox}     
\usepackage{diagbox}       

\usepackage{pgfplots}
\pgfplotsset{compat=1.18}

\setcopyright{acmlicensed}
\copyrightyear{2018}
\acmYear{2018}
\acmDOI{XXXXXXX.XXXXXXX}

\begin{document}

\title[Dynamic Retriever and Generator Selection for RAG]{
One Size Does Not Fit All!\\ Dynamic Retriever and Generator Selection for RAG
}

\author{Neeraj Anand}
\orcid{}
\affiliation{%
  \institution{Media and Data Science Research Lab}
  \city{Adobe Systems}
  \country{Noida, India}
}
\email{neeraja@adobe.com}

\author{Payel Santra}
\orcid{0009-0005-5721-248X}
\affiliation{%
\institution{IACS}
  \city{Kolkata}
  \country{India}
}
\email{payel.iacs@gmail.com}

\author{Partha Basuchowdhuri}
\orcid{0000-0001-7655-7591}
\affiliation{%
\institution{IACS}
  \city{Kolkata}
  \country{India}
}
\email{partha.basuchowdhuri@iacs.res.in}

\author{Debasis Ganguly}
\orcid{0000-0003-0050-7138}
\affiliation{%
  \institution{University of Glasgow}
  \city{Glasgow}
  \country{United Kingdom}
}
\email{debasis.ganguly@glasgow.ac.uk}

\author{Sumit Bhatia}
\orcid{0000-0002-8146-4100}
\affiliation{%
  \institution{Media and Data Science Research Lab}
  \city{Adobe Systems}
  \country{Noida, India}
}
\email{sumit.bhatia@adobe.com}

\begin{abstract}
Retrieval-Augmented Generation (RAG) systems typically employ fixed retriever and generator configurations across queries, despite substantial differences in query complexity and information needs, leading to inefficient allocation of computational resources. While retrieval and generation adaptivity have been studied independently, their joint effect on end-to-end RAG performance remains underexplored. We systematically analyze how retriever and generator
complexity interacts across factoid and multi-hop question answering (QA), including bridge and composition reasoning tasks. Our analysis shows that stronger retrieval generally yields larger gains than increased generation effort, but both exhibit diminishing and non-monotonic returns, indicating that higher-complexity configurations are not uniformly better across queries.

Motivated by these findings, we introduce DRAG, a query-adaptive framework for selecting retriever-generator configurations. We first propose \qpprag, a training-free routing approach that uses Query Performance Prediction (QPP) signals to guide retriever selection and perplexity-based measures over retrieved context to guide generator selection. We further introduce \sft, a supervised routing approach that fine-tunes an LLM to jointly predict retriever-generator configurations. Across three LLM families and four QA benchmarks, \qpprag~achieves performance comparable to strong static RAG baselines while substantially reducing inference latency, whereas \sft~consistently improves effectiveness over static and training-free adaptive baselines. Overall, DRAG demonstrates that jointly adapting retrieval and generation achieves a more favorable effectiveness-efficiency trade-off than static RAG pipelines.
\end{abstract}


\received{20 February 2007}
\received[revised]{12 March 2009}
\received[accepted]{5 June 2009}

\maketitle


\section{Introduction}

Large Language Models (LLMs) have demonstrated strong capabilities in text generation; however, their responses remain constrained by the knowledge encoded in their parameters during pretraining~\cite{lewis2020retrieval}. As a result, LLMs often struggle to reliably generate or reason over information that is absent from, outdated 
in, or only sparsely represented in their training data.
Retrieval-Augmented Generation (RAG) addresses this limitation by explicitly incorporating external knowledge sources into the generation process \cite{lewis2020retrieval,karpukhin2020dense,ye-etal-2023-fid}. A typical RAG pipeline consists of two components: (i) a \textit{retriever} (such as BM25~\cite{robertson2004understanding}, or a dense ranker such as E5~\cite{wang2022text}) that selects contextually relevant documents from an external corpus, and (ii) a \textit{generator}, which is typically a decoder-based LLM (e.g., LLaMA~\cite{touvron2023llama} or GPT~\cite{dale2021gpt}), which conditions on both the input query and the retrieved contexts to produce a task-specific response. Most existing RAG systems rely on a fixed retriever–generator configuration that is applied uniformly across all queries~\cite{gao2023retrieval,wang-etal-2024-improving-text,zhu2021retrieving,DBLP:conf/sigir/ParryGC24}, thus leading to computational inefficiency and suboptimal effectiveness. However, prior work in IR has long established that queries differ substantially in the complexity of their underlying information needs, which in turn affects retrieval difficulty and downstream generation quality \cite{kanoulas2011evaluating,DBLP:conf/wsdm/DattaGGM22,DBLP:conf/cikm/GangulyLMJ11}.

In a RAG setting, query complexity spans a broad spectrum: from simple factoid queries answerable using a single relevant document~\cite{joshi2017triviaqa,kwiatkowski2019natural}, to compositional multi-hop queries requiring reasoning across multiple evidence sources~\cite{yang2018hotpotqa,trivedi2022musique}, as well as queries involving temporal reasoning~\cite{tempQues}, numerical aggregation~\cite{dua-etal-2019-drop}, or multi-constraint satisfaction~\cite{talmor-berant-2018-web}. For simpler queries, standard sparse or dense retrieval models (e.g., sparse such as BM25 or bi-encoders such as E5~\cite{wang2022text}) are often sufficient without requiring computationally expensive re-ranking strategies~\cite{yani2021challenges, gabburo2024measuring}. In contrast, more complex queries require stronger retrieval and generation pipelines, achieved either through enhanced reasoning capabilities in the generator, such as multi-step reasoning~\cite{10.1145/3774896}, query decomposition~\cite{ammann-etal-2025-question}, increasing the \textit{thinking} capability of the generator, structured generation strategies such as chain-of-thought~\cite{kleinman2025e1}, or by improving the retrieval pipeline using tools such as \textit{retrieve-rerank} pipelines employing cross-encoders~\cite{pradeep2021expandomonoduodesignpatterntext,karpukhin2020dense,colbert_sigir20} and LLM-based listwise rankers~\cite{tamber2023scalingdownlittingup,zhuang2022rankt5finetuningt5text,Sinhababu-2024-FSPRP, zhuang2025rank, zhang-etal-2024-two}. However, these improvements often come at the cost of increased computational overhead and latency~\cite{pradeep2023rankzephyr}. This suggests that a \emph{query-adaptive RAG workflow that dynamically allocates retrieval and generation capacity based on the requirements of each query can potentially improve the overall effectiveness--efficiency trade-off.}

While promising, implementing such an adaptive framework presents two primary challenges: (i) \textit{generalizability}, since routing decisions must remain robust across diverse query types and adapt to evolving model configurations without relying on relevance judgments or answer quality signals at inference time; and (ii) the \textit{trade-off between effectiveness and efficiency}, since increased retrieval or reasoning effort improves answer quality only selectively while substantially increasing latency. To address these challenges, we first conduct a \textbf{systematic empirical analysis of retriever-generator configurations of varying complexity} (Section~\ref{sec:config_analysis}), examining how their interactions affect downstream RAG effectiveness and efficiency. This analysis provides insights into when additional computational capacity is beneficial and motivates query-adaptive configuration selection.

\begin{figure}[t]%
\centering
{\includegraphics[width=0.45\textwidth]{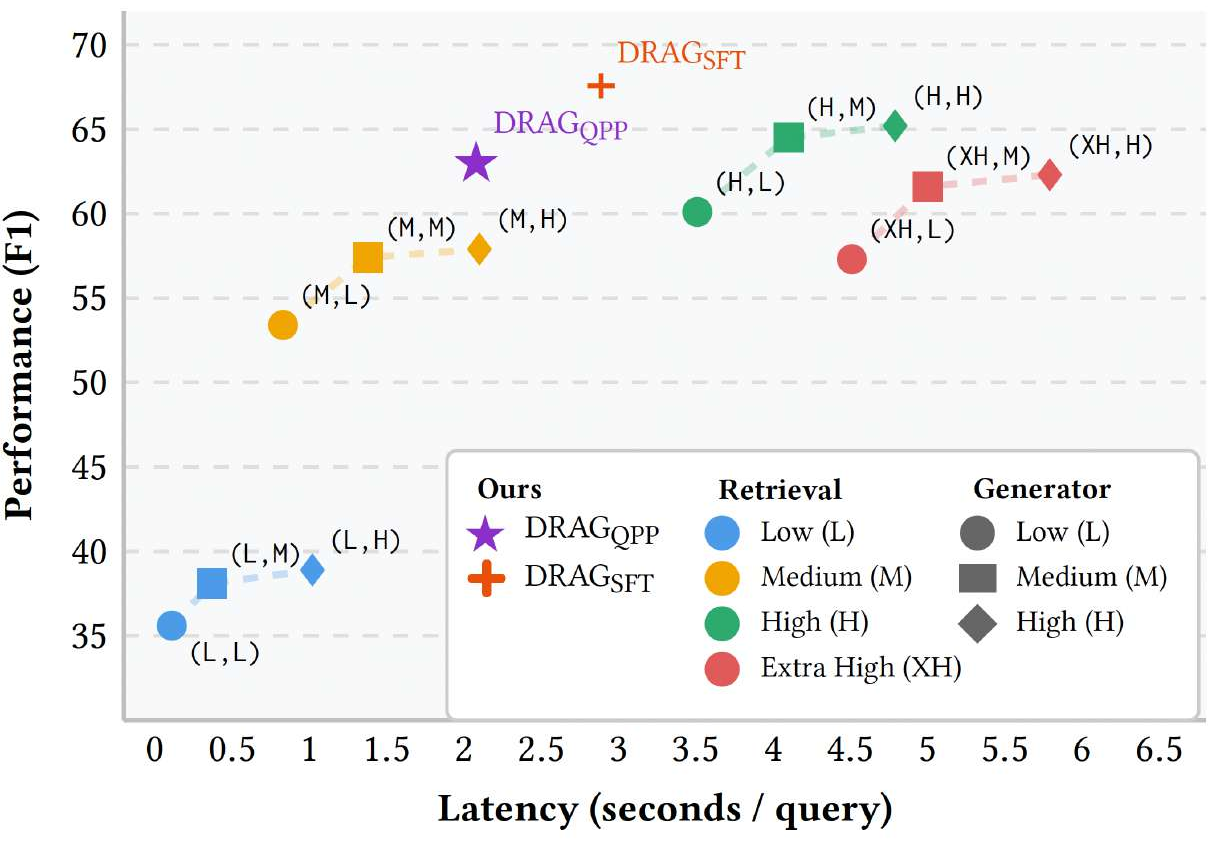}}
\caption{
\small Performance (F1) and efficiency (time/query) on the HotpotQA dataset across $12$ retriever-generator configurations ($4$ retrievers $\times$ $3$ generators) of increasing complexity. The \qpprag\ and \sft\ methods adaptively select query-specific retriever-generator pairs, achieving a favorable effectiveness-efficiency trade-off by attaining strong downstream performance at lower latency than static RAG with high-complexity configurations.
}
\vspace{-3mm}
\label{fig:effect_time}
\end{figure}

To this end, we introduce \textbf{DRAG} (\textbf{D}ynamic \textbf{RAG}), a query-adaptive framework for selecting retriever-generator configurations. We instantiate DRAG through two complementary approaches (Section~\ref{sec:adaptive_framework}): a \emph{training-free} dynamic approach, \textbf{\qpprag}, which routes queries through a cascaded sequence of models of increasing complexity, and a \emph{supervised} approach \textbf{\sft}, that learns to jointly predicts retriever-generator configurations from query signals. Together, these approaches explore the effectiveness-efficiency trade-off (EET) of adaptive RAG under different levels of supervision and computational overhead~\cite{DBLP:conf/sigir/MengAAAR24}.

We study both \qpprag\ and \sft\ under the common \adarag\ framework to analyze their respective strengths and limitations across standard benchmarks as well as evaluate them in out-of-domain (OOD) scenarios. We study OOD robustness using the \textsc{FRAMES} dataset, where routing models are applied without dataset-specific tuning and we show that adaptive retriever-generator selection can generalize beyond the training distribution.

The contributions of our work can be summarized as \DG{too many claims; streamline to 3 at most}:
\uls
  \item We present the \DG{strong words; needs to tone down}first comprehensive empirical study analyzing the joint impact of retriever-generator pair selection on end-to-end RAG performance.
  \item \DG{we shouldn't be claiming for the QPP thing}We propose \textbf{\qpprag}, a training-free, query-adaptive framework that leverages unsupervised retrieval performance prediction (QPP-based~\cite{zhou2007query}) and generator performance prediction (GPP-based~\cite{shorinwa2025survey}) to dynamically balance the EET.
\item We introduce \textbf{\sft}, a supervised framework that treats configuration selection as a predictive task, which improves selection effectiveness over \qpprag, albeit at cost of latency.
  \item We conduct extensive evaluations across three diverse LLM families and four benchmarks -- including out-of-domain evaluation on \textsc{FRAMES} -- demonstrating the robustness and generalizability of our adaptive paradigms across full test sets.
\item To facilitate reproducible research, we release our dataset and code at
\url{https://anonymous.4open.science/r/DynamicRAG-2A01/}.
\ule
\section{Related Works}

\para{Retrieval Augmented Generation} 
Retrieval-Augmented Generation (RAG) improves text generation by grounding LLM outputs in external knowledge, thereby mitigating hallucinations and knowledge limitations. Prior work has explored several forms of retrieval augmentation, including query-based retrieval and retrieval-conditioned generation~\cite{guu2020retrieval,shi2023replug}, latent and logit-level integration of retrieved evidence into language modeling~\cite{khandelwal2019generalization,zhong2022training}, retrieval refinement through reranking, active, and iterative retrieval strategies~\cite{glass2022re2g,jiang2023active,zhang2023repocoder}, as well as architectures that improve reasoning over retrieved documents and multi-document fusion~\cite{izacard2020leveraging,Lan2023CopyIA,10.1145/3746252.3760942}. Another line of work explores supervised fine-tuning of LLMs, that retrofit any LLM with retrieval capabilities~\cite{DBLP:conf/iclr/Lin0CSL00KSLZY24,DBLP:conf/icml/WangPM00SC24}, distill reader signal back into the retriever~\cite{DBLP:journals/jmlr/IzacardLLHPSDJRG23}, or fine-tune the generator to reason over retrieved context~\cite{DBLP:conf/iclr/Lin0CSL00KSLZY24,asai2024self}.

\para{Adaptive Retrieval} Existing studies on adaptive RAG focused on binary decisions, i.e., whether to retrieve or not~\cite{mallen2023not}. This method is sufficient for simple queries, but generally fails multi-hop difficult queries. Thus, some studies introduces a method with a variable number of iterative retrieval steps, i.e., it jointly retrieve-and-rerank until sufficient evidence is accumulated for correct prediction\cite{qi2021answering}. Furthermore, SELF-RAG~\cite{jeong-etal-2024-adaptive,asai2024self} introduces training strategy to retrieve, self-critique and refine the outputs. Unlike our method \adarag, it does not explicitly address the problem of selecting among multiple retriever–generator configurations. In contrast, several works explore adaptivity from reasoning point of view, while assuming a fixed retrieval pipeline. Some existing studies varies the number of reasoning steps~\cite{wang2022self} or samples multiple chain-of-thought reasoning paths to dynamically select the generation path for improving robustness~\cite{yao2022react}. Different from previous approaches, this paper jointly analyzes the impact of retriever and generator choices with different capabilities on RAG framework.




\para{Query Performance Prediction} Query performance predictors can be classified into two major categories: pre-retrieval and post-retrieval. Pre-retrieval prediction technique aims to predict query difficulty by analyzing the prior distribution to querying the given collection of documents~\cite{de2000evaluation,he2006query,mothe2005linguistic}. The post-retrieval prediction methodology, on the contrary, relies on statistical approaches on retrieval scores~\cite{shtok2012predicting,cummins2011improved,zhou2007query,tao2014query}, learning to rank features
inspired predictors~\cite{Chifu2018QueryPPA} and robustness of rank-list perturbation based predictors~\cite{uef_kurland_sigir10,roitman2017robust}. Some recent studies explored supervised and neural approaches for QPP by considering both the content and the score features~\cite{DBLP:conf/wsdm/DattaGGM22,10.1145/3209978.3210041}. In this work, we use these pre-retrieval predictors in our downstream pipeline to guide the selection of an appropriate retriever based on query complexity.



\section{RAG Configuration Analysis and Problem Formulation} \label{sec:config_analysis}

We first formalize the adaptive RAG task and analyze the behavior of \emph{static} RAG systems, where a fixed retriever and generator are used uniformly across all queries. This analysis serves two purposes: (i) characterizing the EET across retriever-generator configurations, and (ii) motivating the need for query-adaptive frameworks.

\para{Task Formulation} \label{ss:problem}
Given a query $q$, a set of candidate retrievers $\Theta = \{\theta_i\}_{i=1}^{m}$ and  a set of candidate generators $\Phi = \{\phi_i\}_{i=1}^{n}$, the objective of adaptive RAG is to select a query-specific retriever-generator configuration $(\theta,\phi)\in\Theta\times\Phi$ that balances downstream answer quality and computational cost.
Unlike static RAG, adaptive RAG seeks to select the optimal retriever and generator based on the query's specific information need to maximize effectiveness while minimizing latency.

Since the effectiveness of a retriever-generator pair cannot be directly observed at inference time (i.e., the ground-truth answer $a_q$ is unavailable), we need an estimator, \DG{Present the existing approaches as a special case of this formulation where they have a different form for $f$}
$f:
(q,\theta,\phi) \mapsto \mathbb{R}$, that predicts the likelihood of generating a correct answer for a given query $q$. The objective is then to select the most efficient configuration that maximizes downstream effectiveness while minimizing computational overhead.

\para{Retriever and Generator Configuration Choice} \label{ss:config_space}
The adaptive setup requires a discrete action space of retrievers $\Theta$ and generators $\Phi$, ordered by increasing computational cost and capability. Following the conventional view in IR, we maintain the progression of retriever complexity from \textit{lexical retrieval} to \textit{dense semantic retrieval} and further to sophisticated \textit{retrieve-and-rerank} pipelines employing query decomposition and cross-encoder rerankers. Similarly, generator complexity scales from lightweight decoding with no or minimal reasoning to longer, multi-step reasoning i.e., extending the reasoning effort or thinking modes of LLMs. \emph{Importantly, our proposed framework is entirely agnostic to the specific number of retriever or generator choices, allowing it to scale seamlessly across finer- or coarser-grained complexity modes.}

\DG{As per the formal definition, GPT-oss-low and GPT-oss-medium are different generators; it's better to stick to that instead of mixing up model families with thinking mode configurations; I don't quite like the idea of referring to specific models and thinking configutation types in this section because a proposed method should be independent of the particular choices; u should state things generally here and move the specific details used in our experiments to the setup section.}

In our experiments for GPT-OSS base-model, we consider four retrieval levels, i.e., \emph{low} (L; sparse retriever), \emph{medium} (M; dense retriever), \emph{high} (H; retrieve-and-rerank pipeline), and \emph{extra-high} (XH; query decomposition with retrieve-and-rerank pipeline), reflecting complex retrieval pipelines with increasing latency and three cascaded levels for retriever component, i.e., \emph{low} (L), \emph{medium} (M) and \emph{high} (H) reflecting the mode of thinking, which means total $12$ configurations  ($4\times3$). Similarly, for Qwen and Gemma, we consider two generator modes depending on thinking availability (thinking-off/on), resulting in $8$ configurations ($4\times2$) per model.

\begin{table}[t]
\centering
\small
\begin{adjustbox}{width=.85\columnwidth}
\begin{tabular}{@{}lcccc@{}}
\toprule
Gen / Ret & Low (\%) & Medium (\%) & High (\%) & Extra-High (\%) \\
\midrule
Low & 11.2 & 5.6 & 12.5 & 6.9 \\
Medium & 5.1 & 3.8 & 5.7 & 4.5 \\
High & 13.0 & 7.1 & 15.6 & 9.0 \\
\bottomrule
\end{tabular}
\end{adjustbox}
\caption{
\small Distribution (\%) of correctly answered HotpotQA queries under an efficiency-aware oracle RAG system across total 12 configurations, with four retriever levels (L, M, H, XH) and three GPT-OSS generator reasoning levels (L, M, H). The results show that a substantial fraction of queries can be answered using low and medium complexity configurations, indicating that considerable latency reductions are achievable through adaptive framework.
}
\label{fig:motiv_pic}
\end{table}

\para{Efficiency-aware Oracle configuration} \label{sec:oracle_config}
\DG{Isn't the motivation here to construct the training examples? "to understand" is a weak motivation because the "action element" is often missed. Making this distinction here and formally presenting the training set will clarify things}To understand the query-level EET in static RAG frameworks, we define an \emph{efficiency-aware oracle setup} that assigns each query to the least computationally expensive configuration of retriever-generator pair to correctly answer the query $q$. Intuitively, the oracle represents an ideal setup that has access to ground-truth effectiveness and therefore reveals how retrieval and generation effort should ideally be allocated across queries. Formally, given a query $q$, the oracle selects 
\begin{equation}
(\theta^*, \phi^*) = \argmin_{\theta,\phi \in \Theta \times \Phi} \pazocal{L}(\theta,\phi) \quad
\text{s.t.} \quad \pazocal{P}\bigl(a_q,\phi(q, \theta_k(q))\bigr) \ge \delta,
\label{eq:oracle}
\end{equation}

where i) $\theta_k(q)$ denotes the top-$k$ documents retrieved by the ranker $\theta$, ii) $\phi \in \Phi: (q, \theta_q) \mapsto \hat{a}$ denotes a generator which takes $q$ along with its retrieved documents $\theta_q$ as input to yield a predicted answer $\hat{a_q}$, iii)
$a_q$ is the ground-truth answer for query $q$, iv) $\pazocal{P}: a_q \times \hat{a_q} \mapsto \mathbb{R}$ denotes downstream effectiveness (measured by metrics such as Exact Match (EM) and F1) between $\hat{a_q}$ and $a_q$, v) $\pazocal{L}: \theta \times \phi \mapsto \mathbb{R}$ is the total latency and vi) $\delta$ represents a effectiveness threshold. When multiple configurations satisfy the threshold, the oracle selects the least expensive one.

In practice, if no configuration satisfies the target criterion (e.g., EM$=1$), we map to the least expensive configuration achieving high partial correctness (F1$\geq0.8$). \DG{what's the basis of choosing these target criteria? can't we present this in a formal way with a particular notation for the evaluation function on an input? something like a $\mu(q)$?}
Following prior works that applied thresholds on answer-level F1 to account for minor lexical variations in partial correct responses~\cite{asai2024self,jiang2023active}, we use $0.8$ as a threshold \DG{did those papers use an identical threshold?}. If no configuration satisfies this criterion, the query is assigned to the highest-complexity setting $(\theta_{XH},\phi_H)$ as a safe fallback.

\begin{figure*}[t]
    \centering
\includegraphics[width=0.96\textwidth]{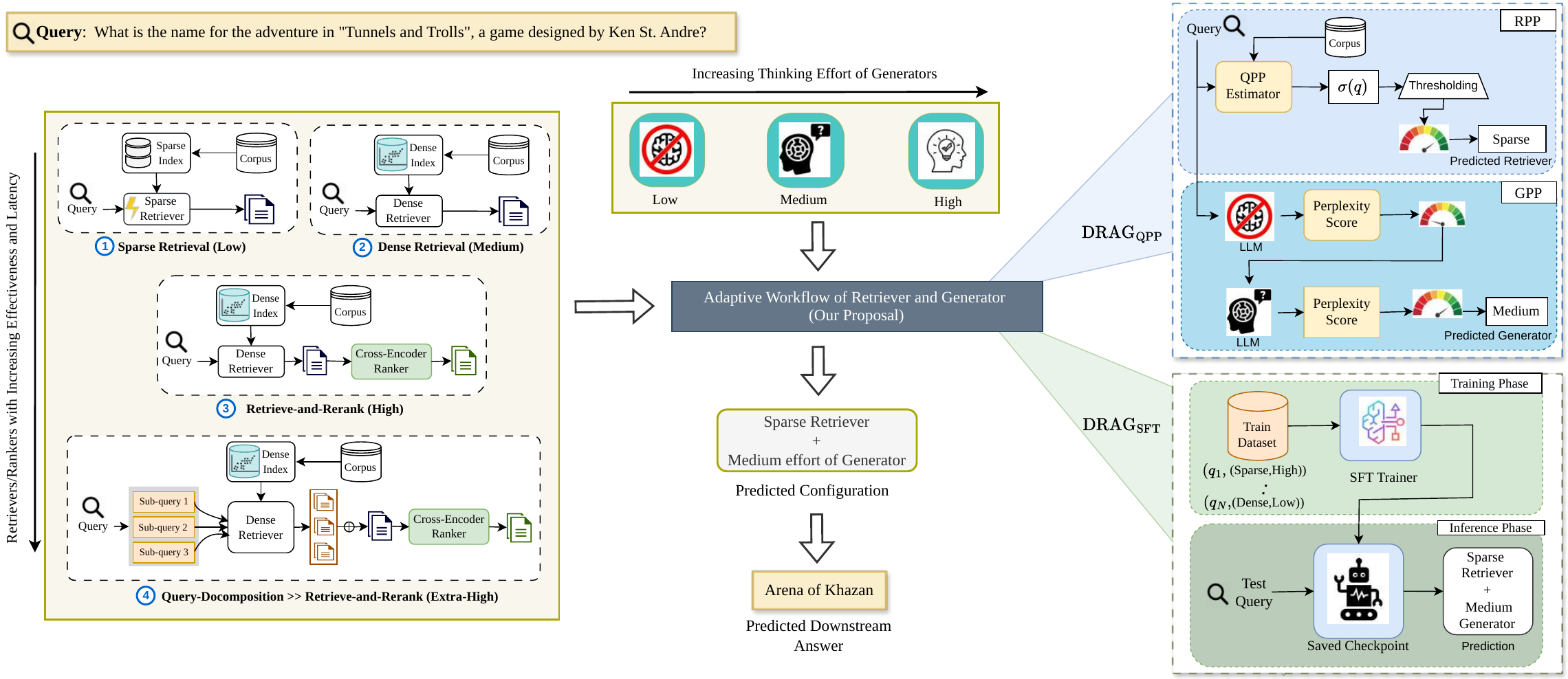}
    \caption{\small Overview of the proposed query-adaptive framework (\adarag) for GPT-OSS. The left panel shows retriever and generator configurations of progressively increasing complexity, ranging from sparse to query decomposition with retrieve-rerank pipelines, and from low to high thinking effort. Given an input query, \adarag\ dynamically selects a query-specific retriever-generator configuration instead of uniformly invoking a fixed pipeline. The top-right panel illustrates the training-free \qpprag\ workflow with 2 modules, \textit{retrieval performance prediction} (RPP) followed by \textit{generator performance prediction} (GPP). The bottom-right panel shows \sft, a supervised framework that learns retriever--generator selection from query--configuration supervision and predicts the optimal configuration during inference. The selected configuration is then executed to produce the downstream answer. While illustrated using GPT-OSS with four retrievers and three generator levels ($4\times3=12$ configurations), the framework naturally generalizes to other model families (e.g., Qwen and Gemma with $4\times2$ configurations).
  For our proposed adaptive \qpprag~approach, we used training-free RPP and GPP-based methods as explained in Section~\ref{sec:adaptive_framework}. \DG{Reduce the white spaces in the figure, e.g., the right column can move closer to the central block; there should be a zoom in indicator to show that the query example at the top is connected to (a specific instance of) the actual query input in the figure; why sub-queries? You didn't mention any of that in the Section 3 and 4; it was all about retriever-generator choices! I don't think the example query and the answer adds any value to the diagram. Training has to be shown separately from inference; Remove the QPP based thing from here; that's a baseline; not our proposed approach}}
        \label{fig:main_fig}
\end{figure*}

\para{Discussion} \label{ss:motivation}
Table~\ref{fig:motiv_pic} reports the oracle distribution across all retriever-generator configurations on the HotpotQA dataset when GPT-OSS is the base model. Note that a substantial fraction of queries are correctly answered using \DG{again in the general sense, i'd just call these generator functions}low- and medium-complexity configurations, indicating that uniformly invoking the most expensive retrieval and reasoning pipeline is often unnecessary.

This data underscores a fundamental reality of RAG optimization: query complexity dictates architectural requirements, meaning some queries are adequately resolved by lightweight configurations while others strictly demand advanced retrieval and reasoning pipelines. For example, a simple factoid question from HotpotQA, \textit{``Alfie Allen played Theon Greyjoy on which show?''} is correctly answered across \textit{all} retriever-generator configurations; for such simple queries, employing expensive retrieval and reasoning pipelines incurs unnecessary computational cost without yielding any performance gains. Conversely, the oracle distribution (\DG{I'd move all results in the "Evaluation" section; there's a lack of continuity in moving to the analysis here; the results don't affect what else we're going to say in this section, right? so why not keep all results in a single section (u can have a separate subsection for the training set analysis)}Table~\ref{fig:motiv_pic}) also reveals that there are a substantial number of queries that are correctly answered only under an intermediate retrieval setting paired with complex reasoner (e.g., $\theta_M$ with $\phi_H$), while other modes of retrievers fail irrespective of generator choice. For instance, the query \textit{``Belle Gold is a fictional character portrayed by an actress of what nationality?''} requires resolving a semantic entity relationship (\textit{Belle Gold} $\rightarrow$ \textit{portraying actress} $\rightarrow$ \textit{nationality}), where dense retrieval ($\theta_M$) successfully retrieves semantically relevant evidence, whereas other retrievers either miss relevant context or retrieved noisy evidence. 
Thus, higher-complexity configurations are not strict supersets of lower ones, and we need \textbf{adaptive per-query routing framework} over the retriever-generator space $\Theta \times \Phi$ to better balance downstream effectiveness and inference latency.

\section{Dynamic RAG Workflow (\adarag)} 
\label{sec:adaptive_framework}
Motivated by the observations in Section~\ref{sec:config_analysis}, which demonstrate that no single configuration is optimal across queries, we introduce \adarag, an adaptive framework that approximates the oracle routing policy of Equation~\ref{eq:oracle} when the ground-truth answer
$a_q$ is unavailable at inference time. Depending on the availability of supervision, we introduce two complementary approaches under this paradigm: \DG{It's better to just mention this in the experiment setup section as a baseline choices specific to our experiments; that way people won't be raising too many questions; here just present the classifier method in a general way}(i)~\textbf{\qpprag}, a training-free framework that sequentially selects retrievers and generators using unsupervised performance signals, and (ii)~\textbf{\sft}, a supervised framework that directly learns optimal, query-specific retriever-generator configurations. An overview of the proposed \adarag\ workflow, including both \qpprag\ and \sft\ frameworks, is illustrated in Figure~\ref{fig:main_fig}.

\subsection{\qpprag} \label{ss:qpprag}
As \DG{an}the initial step towards realizing an optimized adaptive framework, we introduce \qpprag, a training-free routing framework that approximates the oracle routing policy using unsupervised techniques. \DG{Is this an MDP then?}Rather than directly predicting a joint configuration, \qpprag\ decomposes the routing process into two sequential prediction stages: (i) Retriever Performance Prediction and (ii) Generator Performance Prediction. The selected retriever and generator are then used for downstream answer generation.

\subsubsection{\textbf{Retriever Performance Prediction (RPP)}} \label{ss:rpp}
\DG{the formalism can be removed from all of this; the idea itself needs to be presented in a very specific manner just to serve as a baseline... do this in the experiment setup section}

The \DG{do u actually mean this word? is it really the "first" step in an MDP? for SFT, it should not be so!}first module in \qpprag\ framework aims to select the most suitable retriever for each query by \DG{estimating?}analyzing retrieval difficulty~\cite{DBLP:conf/ecir/TianGM26} without requiring relevance judgments at inference time. Because adaptive routing must decide required retriever complexity \textit{before} retrieval is executed, we need a \emph{retriever-independent} signal that relies solely the query text and corpus statistics.
Query Performance Prediction (QPP) directly addresses this need by estimating retrieval quality in the absence of relevance judgments~\cite{datta2024deep,ebrahimi2024estimating}. Motivated by this, we leverage a \emph{pre-retrieval} QPP predictor
$\sigma(q): q\mapsto\mathbb{R}$ as a signal to guide per-query retriever selection~\cite{datta2022relative,meng2025query}; the larger values of $\sigma$ indicate queries expected to have relevant documents after retrieval. Intuitively, queries that are easier to retrieve can be handled by lightweight retrievers, whereas harder queries are escalated to progressively stronger retrieval pipelines.

\para{QPP-aware Retriever Routing}

Rather than evaluating the effectiveness of all retrievers for each query, which would incur substantial computational overhead and defeat the purpose of adaptive routing, we adopt a lightweight query-specific routing strategy that sequentially selects the retrievers with increasing complexity. Since retriever selection must be performed without actually performing each retrieval \DG{why?}, we used \emph{pre-retrieval} QPP predictors, which are \emph{retriever-independent} and depend on the query and corpus statistics.

%
Formally, given a query $q$, we compute a QPP score $\sigma(q)$ and select the least complex retriever satisfying
\begin{equation}
    \theta^{*}(q) = \theta_{i}, \quad \text{where } i = \min \left\{ j \;\middle|\; \sigma(q) \ge \tau_{j} \right\},
    \label{eq:qpp_threshold}
\end{equation}
where $\{\tau_{j}\}_{j=1}^{m-1}$ is a set of $m{-}1$ thresholds (where, $\tau_{j} < \tau_{j+1}, \forall j$) dividing the score range into $m$ intervals. A query interval with lower value of $\sigma$ indicate the need of more complex retrievers, while easier queries computer higher $\sigma$ and are handled by simpler, cheaper retrievers \DG{is this also the workflow for the SFT based approach? these two methods seem to work with two different principles; are they even comparable?}.

In our experiments, we employ \emph{Average IDF} (Avg-IDF)~\cite{kwok1996new} as our lightweight, pre-retrieval QPP signal to capture query specificity with respect to the underlying corpus. Queries dominated by rare, highly informative content terms typically receive higher Avg-IDF scores and are inherently easier to retrieve. Conversely, queries with ambiguous or common terms yield lower scores, indicating they would benefit from more advanced retrieval pipelines. We select Avg-IDF aligned with recent literature demonstrating its robustness~\cite{DBLP:journals/corr/abs-2604-22661}, a choice further validated by our initial experiments against alternative pre-retrieval predictors.



\para{Threshold Selection} \DG{the weakest section of the paper; presenting it as a baseline will save us (no one cares abt a baseline)}
We derive the routing thresholds $\{\tau_{i}\}_{i=1}^{m-1}$ from the Avg-IDF score distribution computed over the training set. Concretely, these thresholds are defined via percentile-based partitioning of the predictor values. We select a base percentile $p$ as a tunable hyperparameter and determine subsequent thresholds using uniform offsets, i.e., $\{p, p+\Delta, p+2\Delta, \ldots\}$, where $\Delta$ controls the interval spacing between adjacent complexity levels. The score values corresponding to these specific percentiles are then fixed as the routing thresholds $\{\tau_i\}$, which are applied to the test data to map incoming queries to their respective retriever complexity levels based on their online QPP scores.





\subsubsection{\textbf{Generator Performance Prediction (GPP)}} \label{ss:gpp}
As the second component of our framework, \qpprag\ selects an appropriate generator for each query, conditioned on the specific retriever $\theta^*$ selected in the previous stage. To capture the generator's uncertainty regarding the retrieved evidence prior to answer synthesis, we evaluate the perplexity of the retrieved context sequence itself under the base language model~\cite{shorinwa2025survey}. \DG{do we need all this if we can train a model?}Perplexity is computed as the exponential of the mean negative log-likelihood of the tokens within the retrieved context documents. Notably, a lower perplexity score indicates that the retrieved evidence is highly consistent with the model’s internal language distribution and parametric expectations. Conversely, a high perplexity score signals that the context is noisy, out-of-distribution, or poorly aligned, which drastically escalates the risk of downstream hallucination~\cite{ji2023survey,xu2024theory}. Consequently, queries paired with low-perplexity context are routed to lightweight generators, while queries yielding high-perplexity context are escalated to more complex, reasoning-intensive models capable of handling noisy evidence.
Formally, for a given query $q$ and retrieved set $\theta^*_k(q)$, we compute the generator-specific perplexity score $h(\theta^*_k(q) \mid q; \phi_j)$ and select the generator according to,

\begin{equation}
    \phi^*(q) = \phi_i, \quad \text{where } 
    i = \min \left\{ j \;\middle|\; h(\theta^*_k(q) \mid q; \phi_j) \le \nu_j \right\}.
    \label{eq:preplexity}
\end{equation}


\para{Threshold on Posterior}
For each generator $\phi_i$, we estimate the distribution of perplexity scores over training queries and define a routing threshold $\nu_i$ as the score corresponding to the $p$-th percentile ($p \in \{10,20,\ldots\}$), where $p$ is treated as a hyperparameter selected on train data. During inference, we select generators sequentially from low to high complexity. We compute the perplexity of the retrieved context and compare it against the corresponding threshold $\nu_i$. If the perplexity exceeds $\nu_i$ -- implying that the query can be answered by the current generation tier's capacity to process the retrieved context's informational density -- the current generator is selected. Otherwise, the query is escalated to the next higher-complexity tier. This process continues until a generator satisfies this threshold criterion or the highest-complexity model is deployed by default.

\subsection{\sft} \label{ss:sft}
\DG{this section is terse, written casually and seemingly in hurry!}
While \qpprag\ provides a lightweight and training-free mechanism for adaptive routing using query and corpus derived signals, its effectiveness relies \DG{not a good thing; presenting an algorithm in so much details and then u urself r pointing out to its limitations; why would i buy ur idea then?}heavily \DG{and u r using such adverbs!} on data-specific threshold selection and the sequential estimation of generation complexity. This limitation motivates a more generalizable supervised approach that directly models retriever and generator selection jointly within a unified framework.

To this end, \sft\ casts routing as a supervised mapping $\hat{\pazocal{P}}: q \mapsto (\theta, \phi)$ from an input query to its optimal retriever–generator configuration. We instantiate $\hat{\pazocal{P}}$ by fine-tuning Qwen \DG{only the upper layers with LoRA or everything?} as the base model on our constructed training dataset $\pazocal{D}_{\text{oracle}}$ (described in Section~\ref{sec:oracle_config}), where each query is paired with its optimal configuration label. The input context consists of the query, the instruction template, and the available configurations, and the model is trained to produce a structured \DG{yeh, so this isn't really an MDP}JSON output of the form {\small \texttt{\{"retriever": $\theta$, "generator": $\phi$\}}} \DG{u have to indicate that $\theta$ is a token (reflecting a model name)}, specifying the selected configuration.

At inference time, given a query $q$, the fine-tuned model predicts the retriever–generator configuration $(\hat{\theta}, \hat{\phi})$ in a single decoding step. The selected retriever $\hat{\theta}$ retrieves the top-$k$ contexts, which are subsequently passed to the generator $\hat{\phi}$ to produce the final answer $\hat{a}_q$. By jointly predicting both components directly from the query, \sft\ avoids the sequential routing and threshold calibration required by \qpprag, while introducing negligible upfront inference overhead.

\section{Experimental Setup} \label{sec:exp-setup}
Our empirical evaluation is designed to answer the following core research questions about the behavior of RAG pipelines. 
\uls 
\li \textbf{RQ1 (Retriever vs. Generator Capacity):} How do retriever and generator complexity individually contribute to end-to-end RAG performance, and does scaling either component in isolation consistently lead to downstream quality gains?


\li\textbf{RQ2 (Dynamic vs. Static RAG Trade-offs):} Can dynamically selecting the retriever and generator configuration optimize the effectiveness-efficiency trade-off \DG{how do u measure the trade-off? is any given more weightage than the other?} (EET) relative to conventional static, fixed-complexity RAG pipelines?

\li\textbf{RQ3 (Framework Generalization):} How effectively do our proposed adaptive routing strategies generalize when evaluated across out-of-domain datasets and unseen model configurations?


\ule

\begin{table}[t]
\centering
\small
\begin{adjustbox}{width=.8\columnwidth}
\begin{tabular}{@{}llrrcc@{}}
\toprule
 Dataset & Task & \#Train & \#Test & $\overline{|Q|}$ & Domain \\
\midrule
TriviaQA & Single-hop & 78,785 & 11,313 & 14.2 & Wiki/Web \\
\cmidrule(l){2-6}
HotpotQA & \multirow{3}{*}{Multi-hop} & 90,447 & 7,405 & 17.8 & \multirow{3}{*}{Wiki} \\
MuSiQue & & 19,938 & 2,417 & 26.4 &  \\
FRAMES & & - & 824 & 16.0 &  \\
\bottomrule
\end{tabular}
\end{adjustbox}
\caption{\small Characteristics of the datasets used in our experiments. We used the entire test set for all evaluations. 
FRAMES is used as an OOD benchmark while the remaining datasets are in-domain.
}
\label{tab:dataset_stats}
\vspace{-3mm}
\end{table}

\subsection{Datasets} 
We evaluate our framework on four question answering (QA) benchmarks selected to represent a spectrum of retrieval and reasoning complexities. \textbf{TriviaQA}~\cite{joshi2017triviaqa} consists primarily of open-domain factoid questions that are typically answerable using a single supporting document, serving as a baseline for settings with limited evidence dispersion. To capture increasing retrieval and reasoning difficulty, we consider three multi-hop QA benchmarks: \textbf{HotpotQA}~\cite{yang2018hotpotqa}, which requires bridging information across indirectly connected documents, \textbf{MuSiQue}~\cite{trivedi2022musique}, which enforces stricter \DG{wrt what?} compositional reasoning (2-4 hops) with widely dispersed evidence and, finally we include \textbf{FRAMES}~\cite{krishna2409fact} as our multi-hop OOD evaluation benchmark, since, unlike the previous datasets, it involves more diverse reasoning requirements (e.g., numerical, temporal, and multi-constraint reasoning), often requiring aggregation across a larger set of documents. Thus, FRAMES provides a challenging testbed for evaluating the robustness and generalizability of the proposed adaptive routing methodologies \DG{only FRAMES?}. \DG{use active voice}Dataset statistics are reported in Table~\ref{tab:dataset_stats}.

\subsection{Retrievers Investigated}

The action space \DG{this doesn't make any sense unless u point me back to the relevant equation(s)!} $\Theta$ should span retrieval pipelines that differ significantly in cost and capability, so the routing policy has substantial trade-offs to learn. In this study, the granularity of $|\Theta|$ is itself a design choice that can vary from simple binary routing to more fine-grained adaptive configurations.

In this work, motivated by the evolution of retrievers and rankers in IR, we consider $\Theta$ as a \DG{informal word; ambiguous unless explained well enough}spectrum of retrieval configurations with progressively increasing complexity: lexical retrieval, dense semantic retrieval, retrieve-rerank pipelines employing cross-encoders, and query decomposition followed by reranking. These configurations have a different \DG{don't overuse a word}spectrum of latency and effectiveness. Although we focus on these dimensions, the framework is general and can accommodate other configuration choices. Particularly, we consider the following:
\uls
\li \textbf{Low ($\theta_L$) (BM25)}: We use BM25~\cite{robertson2004understanding}, a widely adopted lexical model that relies on exact term matching. It is highly efficient but has no semantic awareness.

\li \textbf{Medium ($\theta_M$) (E5)}: We employ \DG{shouldn't this start with a caps?} \textit{e5-large-v2}~\cite{wang2022text}, a dense bi-encoder trained using weakly supervised contrastive learning, which captures semantic relationships under vocabulary mismatch and is more computationally expensive than lexical one.

\li \textbf{High ($\theta_H$) ($\text{E5}\%\text{50} >> \text{BGE}$)}: As the next level of complexity, we employ a retrieve-and-rank model, where we use E5 to retrieve the top-$50$ documents, which are then re-ranked against the query using BGE, a cross-encoder model~\cite{luo2024bge}.
\li \textbf{Extra-High ($\theta_{XH}$) ($\text{QD} >> \text{E5}\%\text{50} >> \text{BGE}$)}: To address multi-hop information needs, we implement a \DG{completely confusing; makes comparisons difficult; if i understand this correctly, enabling this mode on means that the model is itself going to generate queries as a part of the reasoning step itself; so this is rather a part of the generator configuration itself; even with this why do u fix the retrievers then? this should be a generator configuration}decompose-and-rerank pipeline. The query is broken into $s$ atomic sub-queries~\cite{ammann-etal-2025-question}. We retrieve the top-$50$ documents for each sub-query and the original query using E5, creating a candidate pool of $(s+1) \times 50$ documents, which are then re-ranked against the original query using BGE~\cite{luo2024bge}. This incurs the highest computational cost but maximizes relevance for complex tasks.
\ule

\subsection{Generators Investigated}
\DG{yes, so this when u can tie the different generators presented in section 3 as different thinking mode configurations (even of the same generator)}
The space of generator configurations should span varying levels of reasoning capability and computational cost, enabling the routing policy to adapt generation effort based on query complexity. We evaluate three open-source LLM families that provide distinct mechanisms for scaling test-time compute~\cite{snell2024scaling}. Each backbone exposes configurable \emph{reasoning-effort} settings that control the extent of internal chain-of-thought reasoning and inference-time computation. Increasing the reasoning effort generally improves performance on compositional and multi-hop queries, albeit at the cost of higher token usage and increased inference latency.

\DG{this should be presented as table}
\uls
\li \textbf{GPT-OSS}~\cite{openai2025gptoss120bgptoss20bmodel} considers the \textit{reasoning effort} parameter with three settings, \textbf{Low} (minimal internal thought, lowest latency), \textbf{Medium} (moderate reasoning budget), and \textbf{High} (extensive reasoning paths, highest cost but boost in downstream performance).
\li \textbf{Qwen-3}~\cite{qwen3technicalreport} considers a binary thinking-mode switch, \textbf{Low} (Thinking-Off) generates a direct response as a standard instruction-tuned LLM, while \textbf{High} (Thinking-On) emits an explicit step-by-step reasoning trace before the final answer, improving multi-hop performance at the cost of a linearly scaled token budget.
\li \textbf{Gemma-4}~\cite{team2024gemma} similar to the previous setup, it also considers binary thinking control with \textbf{Low} (direct response) and \textbf{High} (reasoning-trace-enabled) modes.
\ule




\begin{table*}[t]
\begin{adjustbox}{width=1.0\textwidth}
\small
\begin{tabular}{@{}l ccc ccc ccc cc cc cc cc cc cc@{}}
\toprule
 &\multicolumn{9}{c}{GPT-OSS (20B)} &  \multicolumn{6}{c}{Qwen (8B)} &  \multicolumn{6}{c}{Gemma (26B)} \\
 \cmidrule(r){2-10}
\cmidrule(r){11-16}
\cmidrule(r){17-22}
 & \multicolumn{3}{c}{EM} & \multicolumn{3}{c}{F1} & \multicolumn{3}{c}{Accuracy} & \multicolumn{2}{c}{EM} & \multicolumn{2}{c}{F1} & \multicolumn{2}{c}{Accuracy} & \multicolumn{2}{c}{EM} & \multicolumn{2}{c}{F1} & \multicolumn{2}{c}{Accuracy} \\
\cmidrule(r){2-4}
\cmidrule(r){5-7}
\cmidrule(r){8-10}
\cmidrule(r){11-12}
\cmidrule(r){13-14}
\cmidrule(r){15-16}
\cmidrule(r){17-18}
\cmidrule(r){19-20}
\cmidrule(r){21-22}
 Ret / Gen & \multicolumn{1}{c}{L} & \multicolumn{1}{c}{M} & \multicolumn{1}{c}{H}  & \multicolumn{1}{c}{L} & \multicolumn{1}{c}{M}  & \multicolumn{1}{c}{H}   & \multicolumn{1}{c}{L} & \multicolumn{1}{c}{M}   & \multicolumn{1}{c}{H} & \multicolumn{1}{c}{L} & \multicolumn{1}{c}{H}  & \multicolumn{1}{c}{L} & \multicolumn{1}{c}{H}  & \multicolumn{1}{c}{L} & \multicolumn{1}{c}{H}  & \multicolumn{1}{c}{L} & \multicolumn{1}{c}{H} & \multicolumn{1}{c}{L} & \multicolumn{1}{c}{H}  & \multicolumn{1}{c}{L}  & \multicolumn{1}{c}{H}  \\
 \midrule
\rowcolor{mygray} \multicolumn{22}{c}{TriviaQA} \\
L   & 53.3$^*$$^\dagger$  & 55.1$^*$$^\dagger$ &  \textbf{56.7}$^*$$^\dagger$  & 63.5$^*$$^\dagger$ & 64.7$^*$$^\dagger$ & \textbf{64.8}$^*$$^\dagger$  & 65.4$^*$$^\dagger$ & 66.4$^*$$^\dagger$ &  \textbf{67.4}$^*$$^\dagger$   & \textbf{57.4}$^*$$^\dagger$ & 57.1$^*$$^\dagger$ & 65.5$^*$$^\dagger$  & \textbf{64.9}$^*$$^\dagger$  & \textbf{66.0}$^*$$^\dagger$ & 64.8$^*$$^\dagger$   & 49.6$^*$$^\dagger$ & \textbf{52.1}$^*$$^\dagger$ & 59.1$^*$$^\dagger$ & \textbf{61.4}$^*$$^\dagger$ & 61.6$^*$$^\dagger$ & \textbf{63.4}$^*$$^\dagger$ \\
M    & 59.6$^*$$^\dagger$ &  62.3$^*$$^\dagger$  & \textbf{64.4}$^*$$^\dagger$   & 71.4$^*$$^\dagger$ &  73.4$^*$$^\dagger$ & \textbf{73.7}$^\dagger$    & 74.4$^*$$^\dagger$ & 76.0$^*$$^\dagger$ &  \textbf{77.1}$^\dagger$  &  64.0$^*$$^\dagger$  & \textbf{64.4}$^\dagger$   & \textbf{73.5}$^*$$^\dagger$ & 73.4$^*$$^\dagger$  & \textbf{74.3}$^*$$^\dagger$ & 74.1$^*$$^\dagger$ & 56.3$^*$$^\dagger$ & \textbf{58.8}$^\dagger$ & 66.8$^*$$^\dagger$ & \textbf{69.0}$^\dagger$ & 70.8$^*$$^\dagger$ & 72.4$^*$$^\dagger$ \\
H & \underline{61.5}$^*$$^\dagger$  &  \underline{64.2}$^*$$^\dagger$  & \textbf{\underline{66.1}}$^\dagger$ & \underline{73.3}$^*$$^\dagger$ &   \underline{75.4}$^*$$^\dagger$ & \textbf{\underline{75.7}}$^*$$^\dagger$ & \underline{76.3}$^*$$^\dagger$ & \underline{77.9}$^*$$^\dagger$ & \textbf{\underline{78.4}}$^*$ & \underline{66.1}$^*$$^\dagger$ & \textbf{\underline{66.8}}$^*$ & \underline{75.7}$^*$$^\dagger$ & \textbf{\underline{75.9}}$^*$$^\dagger$ & \underline{76.5}$^*$ & \textbf{\underline{76.5}}$^*$ &  \underline{58.6}$^\dagger$ & \textbf{\underline{60.1}}$^*$$^\dagger$ & \underline{69.3}$^*$$^\dagger$ & \textbf{\underline{71.1}}$^*$ & \underline{73.9}$^*$$^\dagger$ & \textbf{\underline{76.2}}$^*$ \\
XH  & 60.8$^*$$^\dagger$ &  63.2$^*$$^\dagger$ & \textbf{65.2}$^\dagger$ &  72.6$^*$$^\dagger$  & 74.3$^\dagger$  & \textbf{74.8}$^*$$^\dagger$  & 75.6$^*$$^\dagger$ & 77.1$^*$$^\dagger$ &   \textbf{77.9}$^*$$^\dagger$    & 64.9$^\dagger$  & \textbf{65.6}$^*$$^\dagger$ & \textbf{74.3}$^\dagger$ &  74.1$^*$$^\dagger$ &   \textbf{75.4}$^\dagger$ &   74.7$^*$$^\dagger$ & 57.6$^*$$^\dagger$ & \textbf{59.7}$^*$$^\dagger$ & 68.3$^\dagger$ & \textbf{70.5}$^*$$^\dagger$ & 72.6$^*$$^\dagger$ & \textbf{74.1}$^\dagger$\\
\cmidrule{2-22}
 \qpprag &\multicolumn{3}{c}{65.4} & \multicolumn{3}{c}{73.9} & \multicolumn{3}{c}{76.7} &\multicolumn{2}{c}{64.9} & \multicolumn{2}{c}{74.3} & \multicolumn{2}{c}{75.1}& \multicolumn{2}{c}{58.5} & \multicolumn{2}{c}{68.4} & \multicolumn{2}{c}{74.3}\\
  \sft &\multicolumn{3}{c}{\textbf{68.1}} & \multicolumn{3}{c}{\textbf{77.1}} & \multicolumn{3}{c}{\textbf{79.6}} &\multicolumn{2}{c}{\textbf{67.1}} & \multicolumn{2}{c}{\textbf{76.2}} & \multicolumn{2}{c}{\textbf{76.9}}& \multicolumn{2}{c}{\textbf{61.2}} & \multicolumn{2}{c}{\textbf{71.9}} & \multicolumn{2}{c}{\textbf{76.8}}\\
\rowcolor{mygray} \multicolumn{22}{c}{HotpotQA} \\
L  & 25.5$^*$$^\dagger$ & 28.8$^*$$^\dagger$    & \textbf{30.2}$^*$$^\dagger$ & 35.6$^*$$^\dagger$ &  38.1$^*$$^\dagger$  & \textbf{38.9}$^*$$^\dagger$  & 35.8$^*$$^\dagger$  & 36.4$^*$$^\dagger$ &     \textbf{37.4}$^*$$^\dagger$  &  26.7$^*$$^\dagger$ &  \textbf{27.8}$^*$$^\dagger$ & \textbf{35.5}$^*$$^\dagger$ & 35.4$^*$$^\dagger$  & 33.4$^*$$^\dagger$ & \textbf{33.5}$^*$$^\dagger$ &  25.2$^*$$^\dagger$ & \textbf{26.8}$^*$$^\dagger$ & 33.1$^*$$^\dagger$ & \textbf{34.4}$^*$$^\dagger$ & 33.1$^*$$^\dagger$ & \textbf{33.9}$^*$$^\dagger$  \\
M  & 38.5$^*$$^\dagger$ &   43.9$^*$$^\dagger$   & \textbf{45.5}$^*$$^\dagger$   & 53.4$^*$$^\dagger$  & 57.4$^*$$^\dagger$     & \textbf{57.9}$^*$$^\dagger$  & 53.0$^*$$^\dagger$ &   54.2$^*$$^\dagger$   & \textbf{54.7}$^*$$^\dagger$  & 41.49$^*$$^\dagger$  &  \textbf{45.4}$^*$$^\dagger$ & 53.3$^*$$^\dagger$ & \textbf{57.0} $^*$$^\dagger$ &  49.5$^*$$^\dagger$ &  \textbf{52.3}$^*$$^\dagger$ & 41.6$^*$$^\dagger$ & \textbf{44.3}$^*$$^\dagger$ & 53.5$^*$$^\dagger$ & \textbf{56.1}$^*$$^\dagger$ & 52.3$^*$$^\dagger$ & \textbf{54.0}$^*$$^\dagger$ \\
H & \underline{44.0}$^*$$^\dagger$  &  \underline{49.9}$^\dagger$  & \textbf{\underline{51.6}}$^*$$^\dagger$ & \underline{60.1}$^*$$^\dagger$ &   \underline{64.5}$^*$$^\dagger$ & \textbf{\underline{65.2}}$^\dagger$ & \underline{59.5}$^*$$^\dagger$ & \underline{60.9}$^\dagger$ & \textbf{\underline{61.5}}$^\dagger$ & \underline{46.4}$^*$$^\dagger$ & \textbf{\underline{50.9}}$^*$ & \underline{59.6}$^*$$^\dagger$ & \textbf{\underline{63.9}}$^*$ & \underline{55.2}$^\dagger$ & \textbf{\underline{58.7}}$^*$ &  \underline{46.9}$^*$$^\dagger$ & \textbf{\underline{50.5}}$^*$$^\dagger$ & \underline{60.4}$^*$$^\dagger$ & \textbf{\underline{63.8}}$^*$$^\dagger$ & \underline{59.2}$^\dagger$ & \textbf{\underline{61.8}}$^*$ \\
XH & 41.2$^*$$^\dagger$  &  47.4$^*$$^\dagger$  & \textbf{48.8}$^*$$^\dagger$ & 57.3$^*$$^\dagger$ & 61.6$^*$$^\dagger$ & \textbf{62.3}$^*$$^\dagger$  & 57.0$^*$$^\dagger$ & 57.7$^*$$^\dagger$ & \textbf{58.6}$^*$$^\dagger$  & 43.9$^*$$^\dagger$  &  \textbf{48.1}$^\dagger$ & 56.5$^*$$^\dagger$ &  \textbf{60.3}$^*$$^\dagger$  & 52.5$^*$$^\dagger$ & \textbf{53.3}$^*$$^\dagger$  & 44.6$^*$$^\dagger$ & \textbf{50.5}$^*$$^\dagger$ & 57.5$^*$$^\dagger$ & \textbf{62.7}$^\dagger$ & 56.6$^*$$^\dagger$ & \textbf{60.1}$^*$$^\dagger$ \\
\cmidrule{2-22}
 \qpprag &\multicolumn{3}{c}{49.2} & \multicolumn{3}{c}{63.1} & \multicolumn{3}{c}{60.8} &\multicolumn{2}{c}{48.2} & \multicolumn{2}{c}{61.3} & \multicolumn{2}{c}{55.9}& \multicolumn{2}{c}{48.4} & \multicolumn{2}{c}{61.2} & \multicolumn{2}{c}{59.4}\\
  \sft &\multicolumn{3}{c}{\textbf{53.2}} & \multicolumn{3}{c}{\textbf{67.6}} & \multicolumn{3}{c}{\textbf{63.8}} &\multicolumn{2}{c}{\textbf{50.5}} & \multicolumn{2}{c}{\textbf{63.9}} & \multicolumn{2}{c}{\textbf{58.2}}& \multicolumn{2}{c}{\textbf{51.1}} & \multicolumn{2}{c}{\textbf{64.1}} & \multicolumn{2}{c}{\textbf{62.2}}\\
\rowcolor{mygray} \multicolumn{22}{c}{MuSiQue} \\
L   & 7.1$^*$$^\dagger$   & 7.2$^*$$^\dagger$ & \textbf{7.6}$^*$$^\dagger$  & 11.4$^*$$^\dagger$ & 11.4$^*$$^\dagger$  & \textbf{11.7}$^*$$^\dagger$   & 10.1$^*$$^\dagger$   & 10.6$^*$$^\dagger$ & \textbf{12.0}$^*$$^\dagger$ & 3.9$^*$$^\dagger$  & \textbf{5.5}$^*$$^\dagger$ & 7.9$^*$$^\dagger$ & \textbf{9.2}$^*$$^\dagger$   & 6.5$^*$$^\dagger$ & \textbf{8.2}$^*$$^\dagger$    & 4.4$^*$$^\dagger$ & \textbf{4.5}$^*$$^\dagger$ & \textbf{7.6}$^*$$^\dagger$ & 7.3$^*$$^\dagger$ & \textbf{7.5 }$^*$$^\dagger$& 7.1$^*$$^\dagger$\\
M   & 14.3$^*$$^\dagger$ &  17.8$^*$$^\dagger$ & \textbf{18.5}$^*$$^\dagger$ &  22.1$^*$$^\dagger$ & 25.0$^*$$^\dagger$ & \textbf{25.3}$^*$$^\dagger$  & 20.3$^*$$^\dagger$ & 23.3$^*$$^\dagger$ & \textbf{25.8}$^*$$^\dagger$   & 9.6$^*$$^\dagger$  &  \textbf{15.8}$^*$$^\dagger$ & 15.7$^*$$^\dagger$ &  \textbf{22.1}$^*$$^\dagger$  & 13.4$^*$$^\dagger$ &   \textbf{19.8}$^\dagger$  &  11.8$^*$$^\dagger$  & \textbf{14.8}$^*$$^\dagger$ & 17.3$^*$$^\dagger$ & \textbf{20.0}$^*$$^\dagger$ & 16.7$^*$$^\dagger$ & \textbf{20.0}$^*$$^\dagger$ \\
H & \underline{19.5}$^\dagger$ & 21.1$^*$$^\dagger$ & \textbf{\underline{22.8}}$^*$$^\dagger$ & \underline{27.7}$^*$$^\dagger$ & 29.7$^*$$^\dagger$ & \textbf{\underline{30.1}}$^*$$^\dagger$ & \underline{26.0}$^*$$^\dagger$ & \underline{28.4}$^\dagger$ & \textbf{\underline{31.1}}$^*$ & \underline{11.3}$^*$$^\dagger$ & \textbf{\underline{18.9}}$^*$$^\dagger$ & \underline{18.1}$^*$$^\dagger$ & \textbf{\underline{25.4}}$^*$$^\dagger$ & \underline{15.9}$^*$$^\dagger$ & \textbf{\underline{23.7}}$^*$$^\dagger$  & \underline{15.4}$^\dagger$ & \textbf{17.3}$^*$ & \underline{20.9}$^*$$^\dagger$ & \textbf{22.9}$^\dagger$ & \underline{21.0}$^*$$^\dagger$ & \textbf{23.9}$^\dagger$  \\
XH & 18.6$^*$$^\dagger$ & \underline{21.6}$^*$$^\dagger$ & \textbf{22.1}$^*$$^\dagger$ & 26.8$^*$$^\dagger$ &  \textbf{\underline{29.8}}$^\dagger$ & 29.3$^*$$^\dagger$  & 25.1$^*$$^\dagger$ & 28.1$^*$$^\dagger$ & \textbf{30.5}$^*$$^\dagger$  &  11.0$^*$$^\dagger$ & \textbf{17.9}$^*$$^\dagger$  & 17.6$^*$$^\dagger$ & \textbf{24.0}$^*$$^\dagger$   & 15.3$^*$$^\dagger$ &  \textbf{22.2}$^*$$^\dagger$   & 14.3$^*$$^\dagger$ & \textbf{\underline{17.7}}$^*$$^\dagger$ & 19.8$^*$$^\dagger$ & \textbf{\underline{23.2}}$^*$& 19.5$^*$$^\dagger$ & \textbf{\underline{25.0}}$^*$ \\
\cmidrule{2-22}
 \qpprag &\multicolumn{3}{c}{19.5} & \multicolumn{3}{c}{28.4} & \multicolumn{3}{c}{28.4}& \multicolumn{2}{c}{17.1} & \multicolumn{2}{c}{23.3} & \multicolumn{2}{c}{20.8} & \multicolumn{2}{c}{15.9} & \multicolumn{2}{c}{22.8} & \multicolumn{2}{c}{23.0} \\
 \sft &\multicolumn{3}{c}{\textbf{24.5}} & \multicolumn{3}{c}{\textbf{31.9}} & \multicolumn{3}{c}{\textbf{32.3}}& \multicolumn{2}{c}{\textbf{19.3}} & \multicolumn{2}{c}{\textbf{26.2}} & \multicolumn{2}{c}{\textbf{24.4}} & \multicolumn{2}{c}{\textbf{19.2}} & \multicolumn{2}{c}{\textbf{24.6}} & \multicolumn{2}{c}{\textbf{25.4}} \\
\rowcolor{mygray}\multicolumn{22}{c}{FRAMES} \\
L & \textbf{14.2}$^*$$^\dagger$  &  13.2$^*$$^\dagger$ &   12.1$^*$$^\dagger$    & \textbf{22.9}$^*$$^\dagger$  & 20.6$^*$$^\dagger$ & 18.5$^*$$^\dagger$   &  \textbf{18.2}$^*$$^\dagger$  &   17.5$^*$$^\dagger$  & 17.7$^*$$^\dagger$  & 8.1$^*$$^\dagger$ &  \textbf{12.6}$^*$$^\dagger$ & 13.9$^*$$^\dagger$  & \textbf{19.5}$^*$$^\dagger$  &  9.8$^*$$^\dagger$ &  \textbf{13.8}$^*$$^\dagger$  &  6.3$^*$$^\dagger$  & \textbf{9.5}$^*$$^\dagger$ & 11.2$^*$$^\dagger$ & \textbf{15.0}$^*$$^\dagger$ & 8.1$^*$$^\dagger$ & \textbf{11.3}$^*$$^\dagger$ \\
M & \textbf{18.3}$^*$$^\dagger$    & 18.1$^*$$^\dagger$  & 18.2$^*$$^\dagger$   & \textbf{28.4}$^*$$^\dagger$  & 27.2$^*$$^\dagger$ & 27.1$^*$$^\dagger$  &  21.5$^*$$^\dagger$ &   22.2$^*$$^\dagger$   & \textbf{25.1}$^*$$^\dagger$   & 12.6$^*$$^\dagger$  &  \textbf{18.0}$^*$$^\dagger$ & \underline{20.2}$^*$  & \textbf{26.5}$^*$$^\dagger$  &\underline{14.4}$^*$$^\dagger$ & \textbf{18.9}$^\dagger$   &  10.6$^*$$^\dagger$ & \textbf{15.3}$^\dagger$ & 17.9$^*$$^\dagger$ & \textbf{23.3}$^*$$^\dagger$ & 12.6$^*$$^\dagger$ & \textbf{17.6}$^\dagger$  \\
H &  19.4$^\dagger$ & \textbf{19.5}$^\dagger$ & 17.6$^*$$^\dagger$ & 28.6$^\dagger$ & \textbf{28.8}$^\dagger$ & 26.3$^*$$^\dagger$ & 22.9$^*$$^\dagger$ & 23.3$^*$$^\dagger$ & \textbf{23.5}$^*$$^\dagger$ & 
10.7$^*$$^\dagger$& \textbf{17.0}$^\dagger$ & 19.1$^*$ & \textbf{24.9}$^\dagger$ & 12.1$^*$$^\dagger$ & \textbf{18.0}$^\dagger$  & 11.7$^*$$^\dagger$ & \textbf{16.8}$^\dagger$  & 18.4$^*$$^\dagger$ & \textbf{24.0}$^*$$^\dagger$ & \underline{13.8}$^*$$^\dagger$    & \textbf{19.0}$^*$ \\
XH    & \textbf{21.8}$^*$$^\dagger$  & \underline{21.1}$^*$&    \underline{20.4}$^*$$^\dagger$  & \textbf{\underline{32.1}}$^*$    & \underline{30.6}$^*$$^\dagger$   & \underline{29.6}$^\dagger$     &  \textbf{\underline{26.5}}$^*$  &  \underline{25.2}$^*$$^\dagger$    &\underline{26.0}$^*$  & \underline{11.3}$^*$$^\dagger$  &  \textbf{\underline{18.6}}$^*$$^\dagger$ & 19.6$^*$$^\dagger$ &  \textbf{\underline{27.6}}$^*$  & 13.1$^*$$^\dagger$ &  \textbf{\underline{20.3}}$^*$ &  \underline{12.0}$^*$$^\dagger$  & \textbf{\underline{17.8}}$^*$$^\dagger$ & \underline{19.0}$^*$$^\dagger$ & \textbf{\underline{24.3}}$^*$$^\dagger$  & \underline{13.8}$^*$$^\dagger$ & \textbf{\underline{18.3}}$^*$$^\dagger$  \\ 
\cmidrule{2-22}
 \qpprag &\multicolumn{3}{c}{19.3} & \multicolumn{3}{c}{29.5} & \multicolumn{3}{c}{23.9} &\multicolumn{2}{c}{15.8} & \multicolumn{2}{c}{24.8} & \multicolumn{2}{c}{18.4} & \multicolumn{2}{c}{15.3} & \multicolumn{2}{c}{22.9} & \multicolumn{2}{c}{16.3}\\
  \sft  &\multicolumn{3}{c}{\textbf{23.3}} & \multicolumn{3}{c}{\textbf{33.9}} & \multicolumn{3}{c}{\textbf{27.2}} &\multicolumn{2}{c}{\textbf{19.5}} & \multicolumn{2}{c}{\textbf{28.1}} & \multicolumn{2}{c}{\textbf{21.3}} & \multicolumn{2}{c}{\textbf{18.6}} & \multicolumn{2}{c}{\textbf{25.8}} & \multicolumn{2}{c}{\textbf{18.9}}\\
 \bottomrule
\end{tabular}
\end{adjustbox}
\caption{ 
%
\small
\DG{I'd present results for one model in the table and move the rest to appendices.}
End-to-end RAG performance of three LLMs across four QA benchmarks under all retriever-generator configurations. Rows correspond to retriever complexity levels ($\theta \in \{\text{L}, \text{M}, \text{H}, \text{XH}\}$), while columns correspond to generator complexity levels ($\phi \in \{\text{L}, \text{M}, \text{H}\}$ for GPT-OSS and $\phi \in \{\text{L}, \text{H}\}$ for Qwen and Gemma). We report Exact Match (EM), F1, Accuracy, and average per-query inference time. Among static configurations, \textit{bold} numbers denote the best result within each retriever block (best generator for a fixed retriever), while \underline{underlined} numbers denote the best result within each generator block (best retriever for a fixed generator). We additionally bold the best of the two proposed adaptive methods. Asterisks ($^*$) and Dagger ($\dagger$) indicate statistically significant differences compared to \qpprag~and \sft~respectively~($p < 0.05$). All reported results use 5 contextual examples.}
\label{tab:gpt_oss_tab}
\end{table*}

\begin{table*}[t]
\centering
\small
\begin{adjustbox}{width=0.85\textwidth}
\begin{tabular}{@{}ll cccc cccc cccc cccc}
\toprule
& & \multicolumn{4}{c}{TriviaQA}& \multicolumn{4}{c}{HotpotQA} & \multicolumn{4}{c}{MuSiQue} & \multicolumn{4}{c}{FRAMES (OOD)} \\
\cmidrule(r){3-6}
\cmidrule(r){7-10}
\cmidrule(r){11-14}
\cmidrule(r){15-18}
  & Model  & EM & F1 & Acc & t(s) & EM & F1 & Acc & t(s) & EM & F1 & Acc & t(s) & EM & F1 & Acc & t(s) \\
\midrule
 \multirow{6}{*}{\rotatebox{90}{GPT-OSS}} & \sOracle    & \textbf{77.5} & \textbf{86.1} & \textbf{85.5} & 2.56 & \textbf{63.9} & \textbf{76.9} & \textbf{71.9} & 1.83 & \textbf{37.1} & \textbf{44.5} & \textbf{43.3} & 5.36 & \textbf{38.9} & \textbf{52.4} & \textbf{44.3} & 4.89\\
&  \static & 66.1 & 75.7 & 78.4   & 10.73 & 51.6 & 65.2 & 61.5  & 4.36 & 22.8 & 30.1 & 31.1   & 10.2 & 21.8 & 32.1 & 26.5  & 5.14\\
&  \iclnz  & 42.5 & 52.3 & 53.0 & 1.02 & 20.7 & 27.9 & 34.8 & 0.83 & 5.4 & 10.4 & 21.6 & 1.73 & 10.9 & 17.6 & 23.8 & 1.72\\
&   \wOracle  & 67.3 & 76.2 & 78.9 &   5.74 & 52.3 & 67.9 & 64.2 &  2.67 & 24.6 & 32.2 & 32.5 &   6.43 & 22.8 & 33.7 & 27.8 &  3.20 \\
 &   \configZero   & 52.7 & 65.8 & 65.2   & 5.12 & 43.7 & 57.9 & 55.3 &  2.87 & 16.8 & 26.7 & 26.9 &   6.45 & 16.4 & 27.3 & 21.1 & 3.22 \\
 \rowcolor{mygray} 
& \qpprag  & 65.4 & 73.9 & 76.7 &   5.14 & 49.2 & 63.1 & 60.8 &  2.18 & 19.5 & 28.4 & 28.4 &  6.36 & 19.3 & 29.5 & 23.9 &  3.15 \\
\rowcolor{mygray} 
  &   \sft    & \underline{68.1} & \underline{77.1} & \underline{79.6} & 5.37 & \underline{53.2} & \underline{67.6} & \underline{63.8}   & 3.11 & \underline{24.5} & \underline{31.9} & \underline{32.3}   & 6.85 & \underline{23.3} & \underline{33.9} & \underline{27.2} &  3.28 \\
\midrule
 \multirow{6}{*}{\rotatebox{90}{Qwen}} & \sOracle     & \textbf{76.0} & \textbf{84.3} & \textbf{83.3}   & 2.34 & \textbf{61.8} & \textbf{75.3} & \textbf{68.9}  & 1.13 & \textbf{29.6} & \textbf{39.8} & \textbf{35.3}   & 1.65 & \textbf{31.8} & \textbf{45.1} & \textbf{33.7}  & 0.78 \\
&  \static & 66.8 & 75.9 & 76.5 &  8.76 & 50.9 & 63.9 & 58.7 &  3.16 & 18.9 & 25.4 & 23.7 &   8.64 & 18.6 & 27.5 & 20.3 &  4.39 \\ 
&  \iclnz & 50.6 & 58.3 & 56.5  & 0.82 & 21.3 & 29.8 & 24.1 &  0.86 & 4.9 & 11.6 & 7.5 & 1.07 & 10.4 & 17.9 & 11.5 & 1.41 \\  
  &  \wOracle     & 67.2 & 75.6 & 76.3 &   4.37 & 50.8 & 63.9 & 58.9 &  1.28 & 19.9 & 26.5 & 24.9 &   4.87 & 18.8 & 27.8 & 21.1 &  2.30\\
&   \configZero   & 54.7 & 67.8 & 66.5 &  4.35 & 46.3 & 58.4 & 54.4 & 1.18 & 15.3 & 21.6 & 19.2 &  4.76 & 12.7  & 20.5 & 16.8 & 1.93\\
\rowcolor{mygray} 
&  \qpprag  & 64.9 & 74.3 & 75.1  & 4.10 & 48.2 & 61.3 & 55.9 &  1.21 & 17.1 & 23.3 & 20.8  & 4.23 & 15.8 & 24.8 & 18.4 &  1.84\\
\rowcolor{mygray} 
&  \sft    & \underline{67.1} & \underline{76.2} & \underline{76.9} &  4.46 & \underline{50.5} & \underline{63.9} & \underline{58.2}  & 1.37 & \underline{19.3} & \underline{26.2} & \underline{24.4}  & 5.10 & \underline{19.5} & \underline{28.1} & \underline{21.3}  & 2.14 \\
\midrule
 \multirow{6}{*}{\rotatebox{90}{Gemma}} & \sOracle    & \textbf{76.5} & \textbf{85.4} & \textbf{84.2}   & 2.86 & \textbf{63.3} & \textbf{76.7} & \textbf{72.2}  & 1.32 & \textbf{27.7} & \textbf{36.1} & \textbf{36.1}  & 6.87 & \textbf{27.3} & \textbf{38.2} & \textbf{29.8}   & 4.36 \\ 
&  \static  & 60.1 & 71.1 & 76.2   & 10.86 & 50.4 & 63.8 & 61.8  & 4.87 & 17.7 & 23.2 & 24.9  & 11.13 & 17.7 & 24.3 & 18.3   & 5.68 \\ 
& \iclnz  & 55.8 & 64.3 & 66.9   & 1.32 & 27.1 & 36.8 & 30.3   & 1.09 & 9.4 & 18.8 & 11.9   & 1.04 &  16.3 & 26.1 & 18.0  & 1.55\\ 
 &    \wOracle  & 61.4 & 71.7 & 75.8   & 5.7 & 50.8 & 63.4 & 60.5 & 2.7 & 17.4 & 24.4 & 25.1 & 7.7 & 18.2 & 25.5  & 18.6  & 3.4 \\
&   \configZero   & 51.8 & 61.6 & 64.5 &   3.77 & 45.8 & 59.3 & 57.3  & 2.1 & 15.2 & 22.3 & 22.8   & 7.3 & 15.1 & 21.2 &  16.5   & 3.1 \\
\rowcolor{mygray} 
&  \qpprag & 58.5 & 68.4 & 74.3   & 5.34 & 48.4 & 61.2 & 59.4 & 2.2 & 15.9 & 22.8 & 23.0  & 7.16 & 15.3 & 22.9 & 16.3 &  3.35 \\
\rowcolor{mygray} 
&   \sft    & \underline{61.2} & \underline{71.9} & \underline{76.8}  & 5.67 & \underline{51.1} & \underline{64.1} & \underline{62.2} & 2.30 & \underline{19.2} & \underline{24.6} & \underline{25.4} &  7.34 & \underline{18.6} & \underline{25.8} & \underline{18.9}  & 3.51 \\
\bottomrule
\end{tabular}
\end{adjustbox}
\caption{\small We compare static, oracle, and adaptive RAG configurations across \DG{the different datasets (no need to name them)} TriviaQA, HotpotQA, MuSiQue, and FRAMES. Our adaptive methods, \qpprag\ (training-free) and \sft\ (supervised), select retriever–generator from available configurations. We report EM, F1-score, Accuracy, and average inference latency (time/second, represented in table as t(s)). The number of contexts used for few-shot experiments is $5$. All evaluations are conducted on the full test sets to ensure statistical robustness. 
\textit{Bold} denotes the best result among static and adaptive methods, while \underline{underline} indicates the best over our proposed methods (i.e., grey portion). \DG{numbers need to be right aligned; there should be an indication of which ones are oracles (they need to be grouped together and not presented in the middle of baselines)}}
\label{tab:main_results}
\end{table*}



\subsection{Baselines}
We compare our adaptive methodology against the following baselines to evaluate the benefits of dynamic RAG configurations:

\uls 
\li \textbf{\iclnz}: In this baseline, we directly predict the answer to a particular query without any supporting document, serves as lower bound. It shows the impact of generator's parametric knowledge.

\li \textbf{\static}: 
This represents a standard static RAG setup, where a fixed configuration is uniformly applied to all queries, and performance is reported for the best-performing configuration over the dataset.

\li \textbf{\sOracle}: This represents the theoretical upper bound of an ideal adaptive RAG system as mentioned in Equation \ref{eq:oracle} of Section \ref{ss:problem} (Efficiency-aware Oracle Configuration).



\li \textbf{\wOracle}: 
%
Our routing policy of RPP in \qpprag~(in Equation~\ref{eq:qpp_threshold}) is dependent on pre-retrieval QPP, thus we define an upper-bound by replacing QPP with the ground-truth retrieval effectiveness (AP@50) of each retriever. This yields a retrieval-aware, sequential routing strategy that estimates the maximum achievable gain under perfect effectiveness signals.

\li \textbf{\configZero}: In this case, given a query, the generator is provided with a natural language description of the available retriever-generator configurations and is asked to select the most suitable one, relying solely on parametric knowledge rather than retrieved evidence. The chosen configuration is then used to generate the final answer. This shows whether an LM can infer appropriate configurations from the query's information need. This is a basic baseline from \adarag's perspective.
\ule 
\subsection{Experimental Details}
\para{Evaluation Metrics}
We assess the \textbf{effectiveness of the downstream task} using three standard metrics: Exact Match (EM), F1, and \DG{this is LLM Accuracy and needs to be decribed properly (with the prompt and the model used; details can go in the appendix cited from here)}Accuracy (Acc), which capture exact correctness, partial lexical overlap, and answer containment, respectively. We report \textbf{statistical significance} at the query level using McNemar's test for EM and Accuracy (due to comparison between binary outcomes), and a paired $t$-test for F1, which captures the comparison between systems. The \textbf{retrieval quality} is evaluated using \DG{why? are the relevance labels binary? what makes a document "relevant"?}AP@10 to assess the relevance of the retrieved contexts provided to the generator.

\para{Hyperparameter Details}
For Extra-High mode of retriever we use same the target LLM to decompose each query into atomic sub-queries (temperature $=0.1$). In this paper, for RPP module in \qpprag~we employ Avg-IDF as we pre-retrieved QPP which is computed over query terms with respect to the corpus. It uses thresholds set of $75^{\text{th}}$, $50^{\text{th}}$, and $25^{\text{th}}$ percentiles for distinguishing all four retriever levels. For GPP, we use threshold set of perplexity scores at the $20^{\text{th}}$ percentile of the training distribution (where lower is better). Additionally, for \sft~we used \texttt{Qwen3-4B-Instruct}. We train for 5 epochs with a learning rate of 2e-5, a cosine decay schedule with warmup ratio 0.15, weight decay 0.05, and an effective batch size of 256. The maximum sequence length \DG{of what?} is set to 1,280 tokens. The best checkpoint is selected based on combined (retrieval + generator) macro-F1.
For generation, we use GPT-OSS (20B\footnote{\url{https://huggingface.co/openai/gpt-oss-20b}}), Qwen3 (8B\footnote{\url{https://huggingface.co/Qwen/Qwen3-8B}}) and Gemma (26B\footnote{\url{google/gemma-4-26B-A4B-it}}) both deployed using the vLLM inference engine. To accommodate the necessary reasoning tokens required by these strategies, we set the maximum generation budget (\texttt{max\_new\_tokens}) to $512$, $2048$, and $4096$ for the \textbf{Low}, \textbf{Medium}, and \textbf{High} modes, respectively. All experiments are conducted on NVIDIA A100 GPUs with a batch size of $12$.

\section{Results and Discussion}
\label{sec:results}

We present our experimental results to answer the proposed research questions in Section~\ref{sec:exp-setup}. 

\subsection{Effect of Retriever and Generator Complexity}
\label{sec:rq1}
Table~\ref{tab:gpt_oss_tab} reports the performance of each retriever-generator pair in terms of EM, F1-score, and Accuracy across all four QA benchmarks using three different families of LLMs. To address \textbf{RQ1}, we examine how varying the complexity \emph{individually} influences end-to-end RAG performance, and whether scaling either component separately consistently yields gains. 

Overall, we observe that \textbf{increasing component complexity leads to performance improvement, but the gains are not monotonic}. When we fix the choice of generators and increase the retriever complexity, performance generally improves from lexical retrieval ($\theta_L$) to dense ($\theta_M$) and reranking-based pipelines ($\theta_H$), with gains in all datasets, particularly higher gains for multi-hop benchmarks such as HotpotQA, MuSiQue, and FRAMES, where evidence is dispersed across multiple documents and stronger retrieval is needed. However, it saturates at higher retrieval levels (e.g., $\theta_H\rightarrow\theta_{XH}$) for all datasets except FRAMES, i.e., \textbf{diminishing returns from retrieval scaling} despite substantially higher computational cost \DG{which part of the results show this (guide the reader by giving an example from the table)}. A likely reason of the effect is less evident in FRAMES is that its queries impose multiple constraints (numerical, temporal, multi-entity) and require evidence assembled from genuinely distinct sub-aspects, so atomic sub-query retrieval supplies a signal that single-pass retrievers miss.

In contrast, increasing generator reasoning effort under a fixed retriever (particularly, for weak retrieval) shows comparatively smaller and inconsistent gains for all datasets. For example, for multi-hop dataset, MuSiQue, increasing reasoning effort under weak retrieval ($\theta_L$) yields only a \textbf{7\% improvement} in terms of EM, but improves performance by \textbf{16.9\%} in terms of EM for $\theta_H$. It shows that \textbf{reasoning cannot compensate for poor retrieval quality} and \textbf{generator effectiveness is fundamentally constrained by retrieval quality}, with stronger retrieval typically yielding larger gains than increased reasoning effort alone.

\para{Effectiveness--Efficiency Trade-off}
Table~\ref{tab:gpt_oss_tab} reports downstream effectiveness, while Table~\ref{tab:LATENCY} shows per-query latency for static RAG configurations. We observe a clear trade-off between effectiveness and computational cost, i.e., increasing retriever or generator complexity incurs substantially higher latency while providing only marginal downstream gains. For example, on HotpotQA (GPT-OSS), increasing generator complexity from $\phi_M$ to $\phi_H$ under $\theta_H$ improves EM only marginally (49.9$\rightarrow$51.6), while increasing latency (2.34s$\rightarrow$4.36s, per query). Similarly, under $\theta_{XH}$, the same increase very marginally (47.4$\rightarrow$48.8) despite a high latency increase (4.10s$\rightarrow$5.87s, per query). Moreover, stronger configurations are not always strictly better (e.g., $\theta_H$ outperforming $\theta_{XH}$), so \textbf{uniformly allocating maximum retrieval and reasoning effort is inefficient}. These findings motivate an \textbf{adaptive routing strategy} that dynamically allocates retrieval and generation complexity based on query requirements.

\begin{table}[t]
\centering
\small

\begin{minipage}{0.34\columnwidth}
\centering
\textbf{(a) Static RAG Latency}
\vspace{1mm}
\begin{tabular}{lcccc}
\toprule
$\phi \backslash \theta$ & L & M & H & XH \\
\midrule
L & \gbox{0.11} & \gbox{0.33} & \gbox{1.04} & \rbox{3.51} \\
M & \gbox{0.38} & \gbox{0.67} & \gbox{2.34} & \rbox{4.10} \\
H & \gbox{1.02} & \gbox{1.70} & \rbox{4.36} & \rbox{5.87} \\
\bottomrule
\end{tabular}
\end{minipage}
\hfill
\begin{minipage}{0.5\columnwidth}
\centering
\textbf{(b) \adarag~Latency Breakdown}
\vspace{1mm}
\begin{tabular}{lcc}
\toprule
Time & \qpprag & \sft \\
\midrule
Decision & 0.28 & 0.36 \\
RAG & 1.90 & 2.75 \\
\cmidrule(r){2-3}
Total & \gbox{2.18} & 3.11 \\
\bottomrule
\end{tabular}
\end{minipage}

\caption{\small Per-query latency (seconds) on HotpotQA using GPT-OSS. (a) Static retriever-generator configurations across retrieval ($\theta$) and generator ($\phi$) complexity levels. (b) Latency breakdown of adaptive methods, where decision time denotes routing overhead (IDF computation and perplexity estimation) and RAG time corresponds to downstream retrieval and generation. Retrieval indexing is excluded, as it is a one-time offline preprocessing cost. \gbox{Green} and \rbox{red} cells indicate, respectively, lower and higher total latency w.r.t \sft. \DG{the presentation of results an color convention is confusing}}
\label{tab:LATENCY}
\vspace{-2mm}
\end{table}
\subsection{Dynamic vs. Static RAG Performance} 
\label{sec:rq2}
To address RQ2, Table~\ref{tab:main_results} compares our adaptive methods against oracle setup, \sOracle\ and other baselines across downstream QA metrics for three LLM families. Being an oracle setup, \sOracle\ consistently outperforms all adaptive methods and baselines. This performance gap highlights the difficulty of routing under realistic settings, where configuration decisions must be made with or without supervision.
Among the baselines, \configZero\ consistently outperforms \iclnz\ across nearly all datasets (except TriviaQA with Gemma), suggesting that directly prompting an LLM to predict the answer is insufficient, whereas explicitly predicting the retriever-generator configuration provides a more appropriate downstream prediction. Furthermore, \wOracle\ consistently improves over \qpprag\ across datasets, indicating that retrieval quality estimation remains a key bottleneck for unsupervised adaptive routing. Notably, both \sft\ and \wOracle\ outperform the static configuration, \static, thus query-adaptive selection is more effective than uniformly assigning a fixed high-complexity pipeline to all queries. Finally, \sft\ consistently improves over \qpprag, suggesting that jointly learning query-specific retriever-generator interactions is more effective than sequential threshold-based routing, albeit at the cost of supervision.

\begin{figure}[t]%
\centering
{\includegraphics[width=0.45\textwidth]{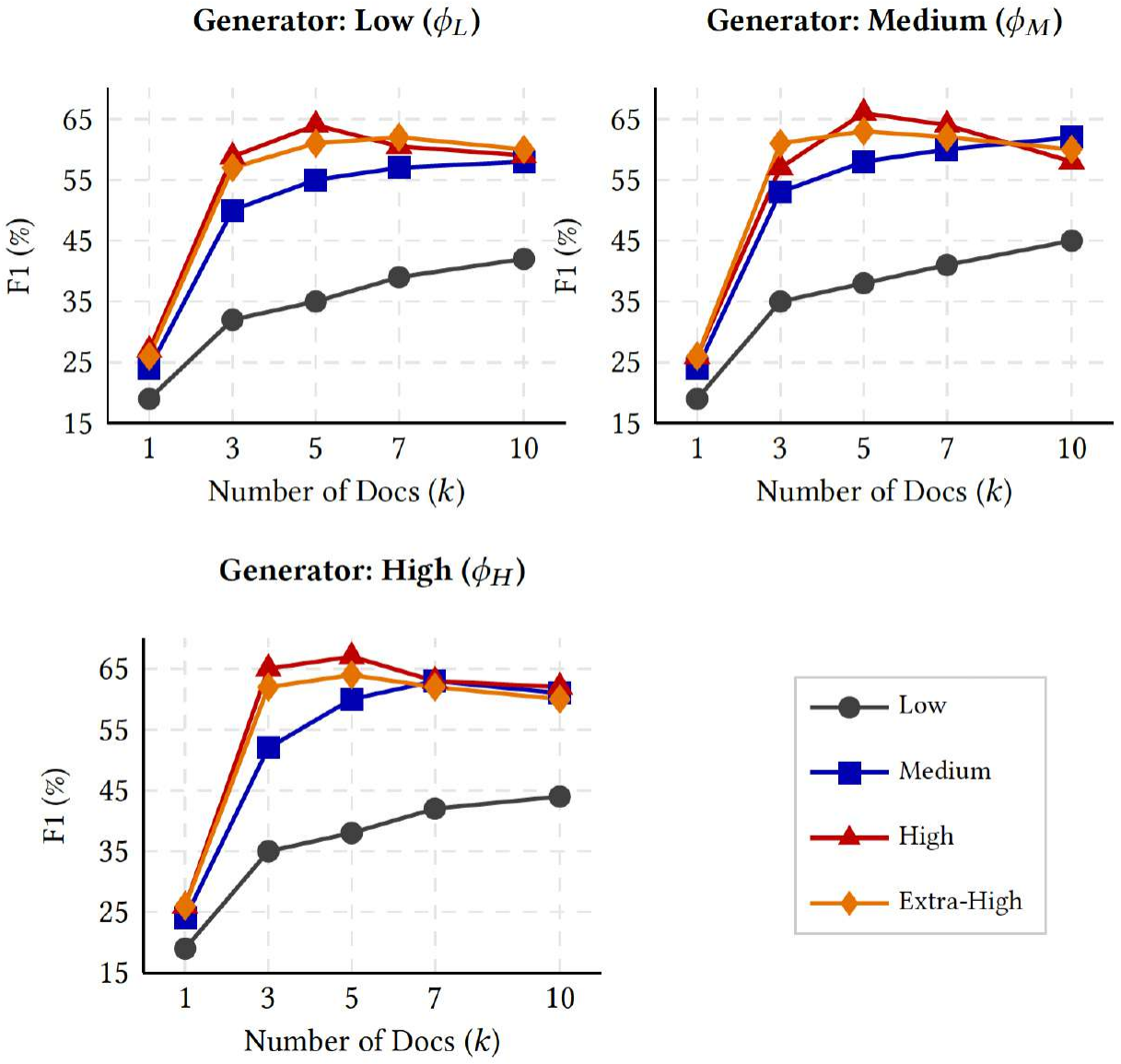}}
\caption{
 \small Effect of change in retrieved document count ($k$) on the performance of HotpotQA dataset. We analyze F1 scores under varying Retriever ($\theta$) and Generator ($\phi$) complexities. The results demonstrate how increasing the number of documents impacts performance across different system configurations. Each line corresponds to a retrieval complexity. The shared legend applies to all subplots.}
\label{fig:gen_complexity}
\end{figure}

\begin{figure}[t]%
\centering
{\includegraphics[width=0.45\textwidth]{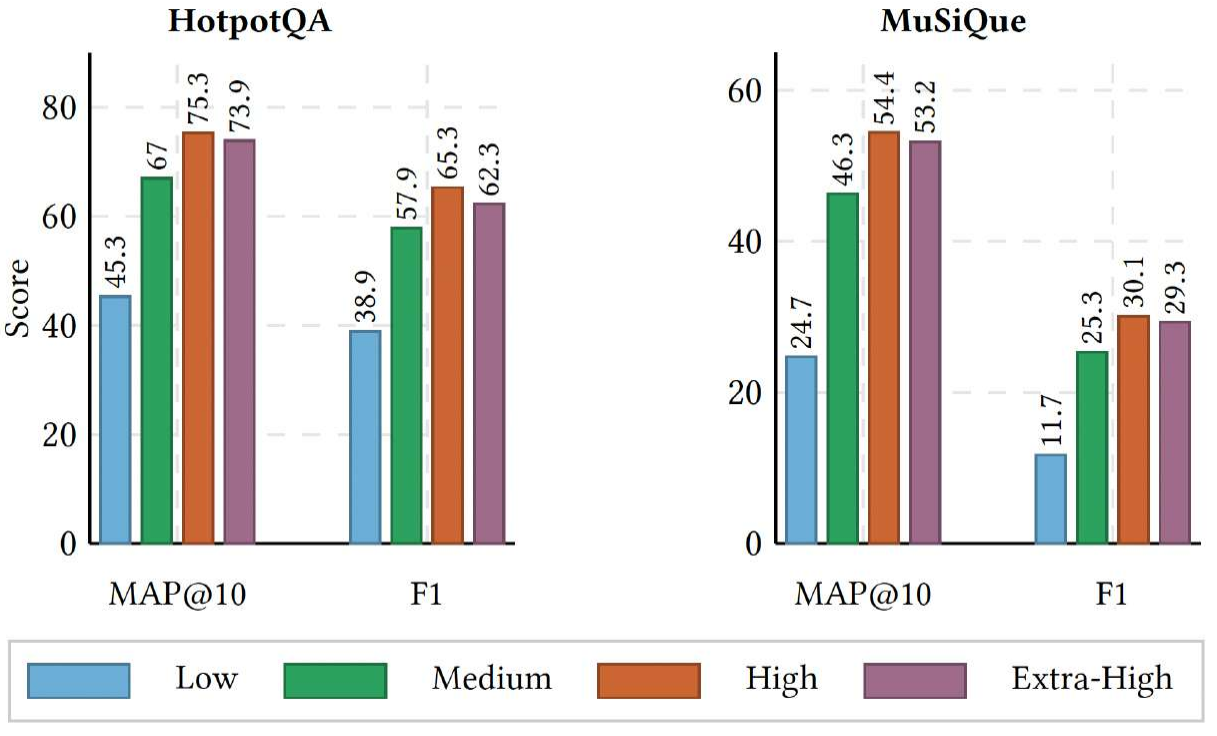}}
\caption{\small Impact of retriever effectiveness on downstream QA performance using GPT-OSS on HotpotQA and MuSiQue. MAP@10 measures retrieval effectiveness, while F1 measures end-to-end QA performance under static RAG configurations. The bars correspond to retriever modes.}
\label{fig:ret_perf}
\end{figure}

\begin{table*}[t]
\centering
\small
\begin{adjustbox}{width=0.87\textwidth}
\begin{tabularx}{\textwidth}{@{}Xlllll@{}}
\toprule
&  \multicolumn{2}{c}{\textbf{\sOracle}} & \multicolumn{2}{c}{\textbf{\sft}} & \\
\cmidrule(l){2-3}
\cmidrule(l){4-5}
\textbf{Query and Gold Answer} & \textbf{Config} & \textbf{Output} & \textbf{Config} & \textbf{Output} & \textbf{Failure Type} \\
\midrule
What is Mohamed Atta's car, manufactured by the company that includes Datsun, an instance of? \textbf{Ans:} Nissan Altima & XH / L & \textit{a 2001 Nissan Altima} & XH / H & \textit{a blue rental car} & Overestimation \\

What is the oldest college in the state that contains Borough House Plantation? \textbf{Ans:} College of Charles Towne & H / H & \textit{College of Charles Towne} & L / M & \textit{Columbia University} & Underestimation \\

What is the tallest building in the state where KZAR is located? \textbf{Ans:} JPMorgan Chase Tower & H / H & \textit{JPMorgan Chase Tower} & L / M & \textit{Transamerica Tower} & Overestimation \\
What is the position of the first governor general of India? \textbf{Ans:} Governor-General of India & L / H & \textit{Governor-General of British India} & L / L & \textit{Warren Hastings} & Wrong Entity \\[8pt]
What team does the person who hit the first home run at the new South Bronx stadium play for? \textbf{Ans:} New York Yankees & M / L & \textit{New York Yankees} & H / H & \textit{Indians} & Wrong Entity \\
\bottomrule
\end{tabularx}
\end{adjustbox}
\caption{\small Qualitative failure cases on MuSiQue dataset. Each row shows a query where \sft\ selects the failure/suboptimal cases relative to the oracle \sOracle, along with the resulting output error. Config columns represent the mode of
\textit{retriever/ generator.
}
\DG{what does overestimaion even mean for a question that seeks an entity? Why should we compare with Oracle? It's just an upper bound. You should compare against best-retriever adaptive-gen (BRAG) and adaptive-ret best-gen (ARBG)} never use vspace; the publishers won't allow this anyway}
\vspace{-3mm}
\label{tab:failure-cases}
\end{table*}

\para{Performance vs Latency} Figure~\ref{fig:effect_time} and Tables~\ref{tab:main_results} illustrate the effectiveness and latency (sec/query) values of adaptive and static RAG systems. 
Overall, increasing retriever and generator complexity improves downstream performance but at substantially higher inference cost, with marginal gains at stronger configurations. We also observe \textbf{a clear trade-off emerges between our two proposed \adarag~methods}, \sft\ consistently achieves better downstream effectiveness than \qpprag, typically improving performance by approximately 2-12$\%$ across datasets, but incurs a modest increase in latency (approximately 5-40$\%$). While \qpprag\ remains training-free and lightweight, \sft\ leverages supervision to model query-specific interactions, resulting in better downstream performance with higher inference cost.

\para{Significance Testing} Table~\ref{tab:gpt_oss_tab} further highlights the statistical behavior of the \adarag\ framework. Across in-domain datasets, \sft\ consistently yields statistically significant gains \DG{in terms of paired t-test?} over static configurations, whereas \qpprag\ often remains statistically comparable to static RAG with medium/high-complexity generators. Under the OOD setting (FRAMES), although \sft\ generally maintains stronger performance, improvements are not always statistically significant, suggesting that transferring routing decisions across datasets with different query types remains challenging.

\para{Latency} Table~\ref{tab:LATENCY} reports the per-query latency breakdown of adaptive RAG systems. We observe that the routing cost, i.e., model configuration prediction cost, is relatively less. Instead, most of the latency difference arises from retrieval and downstream generation ($1.90$s vs.\ $2.75$s RAG time), here \sft\ picks higher-complexity configurations, trading latency for effectiveness, while \qpprag\ favors cheaper settings at the cost of a modest accuracy drop.


\subsection{Model Generalization}
\label{sec:rq3}
To address RQ3, we evaluate the robustness of our adaptive methods under an OOD setting using FRAMES, where models don't rely on dataset-specific training (i.e., use training data of MuSiQue and are evaluated directly on FRAMES). 
From Tables~\ref{tab:main_results} and~\ref{tab:gpt_oss_tab}, we observe that both \qpprag\ and \sft\ provide better results w.r.t all the baselines, including \wOracle~across all three LLM families. Interestingly, Table~\ref{tab:gpt_oss_tab} shows that for GPT-OSS on FRAMES dataset, the best static RAG performance is achieved under low or medium-reasoning generator settings with a fixed retriever choice, thus uniformly increasing reasoning effort is not always beneficial under distribution shift. In contrast, our adaptive methods not only achieve better downstream effectiveness than \static~ and other baselines, but also incur a lower inference cost. Thus, our proposed framework, \adarag\ is robust and generalizable across OOD datasets.



\subsection{Ablation Studies} \label{sec:ablation}

\para{Impact of Increasing Context Size} Figure~\ref{fig:gen_complexity} analyzes the sensitivity of RAG performance to the number of retrieved documents $k$ across retrieval complexity levels and generator reasoning modes. While low-complexity retrieval ($\theta_L$) shows steady improvement with increasing $k$, reflecting its lower per-document precision and reliance on broader context coverage, higher-complexity retrievers ($\theta_M$, $\theta_H$, $\theta_{XH}$) plateau early, as their stronger relevance signals are largely captured within a compact candidate set. This saturation behavior is consistent across all three generator reasoning levels, suggesting it is driven by retrieval quality rather than generation capacity, and collectively justifies our choice of $k = 5$ as a balance between evidence coverage and computational efficiency.

\para{Correlation between retrieval and downstream performance} 
Figure~\ref{fig:ret_perf} examines the relationship between retrieval effectiveness (MAP@10) and downstream QA performance (F1) across retrieval complexity levels on HotpotQA and MuSiQue. A clear positive correlation is observed: as retrieval complexity increases from lexical ($\theta_L$) to dense ($\theta_M$) and reranking-based pipelines ($\theta_H$, $\theta_{XH}$), both MAP@10 and F1 improve consistently and proportionally across both benchmarks, confirming that stronger retrieval directly translates to better answer generation, particularly for multi-hop queries where evidence is dispersed across multiple documents. 

\subsection{Per-Query Analysis}
Table~\ref{tab:failure-cases} presents qualitative failure cases of our best performing adaptive method \sft\ (as shown in Table~\ref{tab:gpt_oss_tab}) on MuSiQue, revealing three primary error patterns. \textit{Overestimation} occurs when \sft\ routes simple queries to unnecessarily expensive configurations (e.g., $\theta_H$/$\phi_{XH}$ instead of $\theta_L$/$\phi_{XH}$), introducing noisy evidence that degrades answer quality. \textit{Underestimation} occurs when \sft\ assigns insufficient capacity to genuinely complex queries (e.g., $\theta_M$/$\phi_L$ instead of $\theta_H$/$\phi_H$), leading to factually incorrect answers. Finally, \textit{wrong entity} errors occur when suboptimal configuration selection (e.g., $\theta_H$/$\phi_H$ instead of $\theta_L$/$\phi_M$) causes the model to retrieve incorrect evidence, leading to entity confusion during multi-hop reasoning. Together, these cases highlight that routing errors are systematically tied to query complexity estimation, and that better query representations could further improve adaptive configuration selection.
\section{Conclusions and Future Work}

We presented a systematic study of how retrieval complexity and generation effort jointly affect end-to-end RAG performance. Our results show that increasing complexity does not consistently improve accuracy, and that high-complexity configurations are not strict supersets of simpler ones. This highlights the non-monotonic, query-dependent nature of RAG and the limitations of fixed retriever–generator settings.
%
%
To address this, we introduced a training-free approach (\qpprag) and a supervised approach (\sft) under a common framework \adarag~for per-query retriever-generator selection.
Across multiple QA benchmarks, \qpprag~matches strong static baselines while reducing latency, and \sft\ further improves the effectiveness–efficiency trade-off, achieving performance close to oracle routing.


Future work will focus on developing more expressive routing policies that better capture the non-separable interactions between retrieval and generation, including richer supervision and reinforcement learning for optimizing cost–quality trade-offs. \DG{adaptive choice of general tools in a tool calling based agentic framework?}

\section*{GenAI Usage Disclosure}
Generative AI tools were not used for core idea generation or experimental design. Its use was limited to minor writing and formatting.


\bibliographystyle{ACM-Reference-Format}
\bibliography{refs}

\end{document}